\documentstyle[secnumtab,epsfig]{fbssuppl}

\def\slash#1{\rlap{$#1$}/} 
\def\lsim{\mathrel{\raise3pt\hbox to 8pt{\raise -6pt\hbox{$\sim$}\hss{$<$}}}}

\title{Dimensional Power Counting in Nuclei}
\author{J. L. Friar\thanks{{\it E-mail address:} 
friar@sue.lanl.gov}}
\institute{Theoretical Division, Los Alamos National Laboratory,
MS B283, Los Alamos, NM 87545, USA}

\begin{document}

\maketitle
\begin{abstract}
Weinberg's dimensional power counting for nuclei is reprised in this primer. 
The role of QCD and chiral symmetry in constructing effective Lagrangians is
discussed. The two-scale hypothesis of Manohar and Georgi is combined with power
counting in amplitudes to shed light on the scales of nuclear kinetic and
potential energies, the size of strong-interaction coupling constants, the size
of meson-exchange currents, and the relative sizes of $N$-nucleon forces. 
Numerous examples are worked out and compared to conventional nuclear models.
\end{abstract}

\section{\underline{Nuclear Perspectives}}

What is a nucleus and what are its constituents?  This question  
has no direct answer, or even a unique one.  Indeed, the proper  
answer depends on the context of the question.  Alternatively, the  
answer depends on the energy scale at which we are probing the  
nucleus.  At extremely high energies such as those of CERN or  
RHIC, we might suppose that the appropriate degrees of freedom  
(d.o.f.) are quarks and gluons.  At very low energies (such as  
nuclear ground states), the more natural degrees of freedom  
would be nucleons and possibly pions.  After all, collisions of a  
nucleus with low-energy projectiles ($\sim$ tens of MeV) eject  
nucleons.  This traditional description of a nucleus as interacting  
nucleons would seem at first glance to have little direct  
connection with QCD, the fundamental theory of the strong  
interactions.  The former description in terms of quarks and  
gluons obviously has such a connection, since the beautiful  
economy and symmetry of the theory is manifest in terms of  
these d.o.f\cite{1,2}.

A close connection between the two descriptions must exist, but  
it is murky.  The traditional picture of nuclei has been very  
successful, provided that the complexities of the short-range  
nuclear force are resolved by appeal to experiment.  While this  
works for the nucleon-nucleon ($NN$) force\cite{3}, we have as yet been nearly  
powerless to resolve the three-nucleon force in this way, although  
substantial progress is being made\cite{4}.  The few-nucleon systems  
are arguably the area of nuclear physics where the most progress  
has been made on seminal problems during the previous decade  
or so\cite{5}.  We expect (and will see) that these systems do provide us  
with substantial information on dynamics, assuming that we can  
interpret it.  That is the purview of this article.  Conversely, the  
scheme that is outlined below has the potential to augment  
greatly our understanding of nuclear dynamics, and particularly  
the role and appreciation of few-nucleon systems (our testing  
ground) in this endeavor.

Traditional nuclear physics is the domain of large coupling  
constants (the interaction is strong), which presupposes that  
perturbation theory {\it cannot} converge.  Although we will see  
below that this is true in part, it is not true {\it in toto}, and this  
caveat makes progress possible.  Connections to QCD (at least  
the qualitative ones) are easy to make and most are long known\cite{1,2},  
and follow from the chiral symmetry (CS) manifested by QCD.\\

\indent $\bullet$ The pion has a very small mass, $m_\pi$.

\indent $\bullet$ The pion is a pseudoscalar particle ($J^\pi = 0^-$).

\indent $\bullet$ The pion is an isovector particle.

\indent $\bullet$ Chiral symmetry forbids large $\pi N$ interactions.

\indent $\bullet$ There is a large-mass scale, $\Lambda \sim$ 1  
GeV, associated with QCD.\\

\noindent The first four items involve the pion and are extremely important, 
but have been known for decades\cite{6}. The last item is new\cite{1,7,8}, and 
its consequences for nuclear physics are more subtle and will shape  
the contents of this article. In the past, the lack of any apparent dynamical 
scales in nuclear physics made progress in interpreting results rather 
difficult.

The (broken) chiral symmetry referred to above is a consequence  
of (nearly) massless quarks in the QCD Lagrangian.  This  
symmetry is reflected in the Goldstone mode by (nearly)  
massless pions\cite{8}.  An analogous chiral symmetry exists in QED if  
the electron mass is set to zero; this symmetry produces a  
vanishing amplitude for (relativistic) electrons scattering  
backwards from the nuclear charge distribution.  The very small  
pion mass plays a significant role in nuclear physics, producing  
the longest-range part of the strong force.  The pseudoscalar  
nature of the pion guarantees a spin-dependent interaction with a  
nucleon, and this leads to the tensor force, the dominant  
component\cite{9} of binding in few-nucleon systems ($\langle V_{\pi} 
\rangle / \langle V \rangle \sim$
75\%) and possibly in all nuclei.  The existence of charged pions  
guarantees them a large role in electromagnetic (EM) meson-exchange  
currents\cite{10}, which we will treat later.  The fourth item on the list is  
critical for a tractable theory of nuclei.  It guarantees that a class  
of many-body forces is weak\cite{11,12} (we will see later that they all are  
weak).  Item five allows us to combine phenomenology with  
principle, and is the organizing element of Chiral Perturbation Theory 
($\chi$PT)\cite{1,2,13,14,15}, which we will treat qualitatively.  It is not  
unreasonable to expect that this subject will become increasingly  
important to nuclear physics in the next decade, and this primer was
motivated by that supposition.

As evidence for these views, Table 1.1 shows the (current)  
results of one of the most important (set of) calculations  
undertaken in nuclear physics\cite{16}.  Using the best available $NN$  
force, a weak three-nucleon force (3NF), and no four-nucleon  
force (4NF), the light nuclei ($A \leq 6$) were calculated with an  
uncertainty $\lsim$ 1\%.  The agreement with experiment must be considered 
very good. Our goal will be to understand these results in simple terms.

\begin{table}[hbt]
\centering
\caption{Calculated and experimental ground-state energies (in MeV) of 
    few-nucleon systems, together with (approximate) dates when they were first 
    accurately solved for ``realistic'' potentials.}

\begin{tabular}{|l||llllll|}
\hline
{$^{\rm A}{\rm X} (J^{\pi})$}&\hspace{0.1in} {$^2{\rm H}(1^+)$} & 
{$^3{\rm H}(\frac{1}{2}^+)$} & 
{$^4{\rm He}(0^+)$} &  {$^5{\rm He}(\frac{3}{2}^-)$} & 
{$^5{\rm He}(\frac{1}{2}^-)$} & {$^6{\rm Li}(1^+)$} \\ \hline \hline
{Solved} &\hspace{0.1in}$\sim$1950 & 1984 & 1987 & 1990 &
 1990 & 1995 \\ \hline
Expt. &\hspace{0.1in} -2.22 & -8.48 & -28.3 & -27.2 & -25.8 & -32.0 \\
Theory &\hspace{0.1in} -2.22 & -8.47(2) & -28.3(1) & -26.5(2) & -25.7(2)
& -32.4(9) \\ \hline
\end{tabular}

\end{table}

We summarize this discussion by stating that\\

\noindent $\bullet$ Chiral Symmetry (manifested in QCD)  
provides an organizing principle for nuclear physics.

\section{\underline{Motivation}}

In 1979, Weinberg\cite{8} suggested that a convenient and simple way  
to reproduce the results of current algebra (and go beyond) was to use a 
nonrenormalizable, phenomenological Lagrangian that manifests  
chiral symmetry.  Such a scheme would produce amplitudes of  
the form: $T \sim$ $E^{\nu}$, where $E$ is the energy.  This was  
obtained using dimensional analysis and specified $\nu$ to be  
an integer determined by the type of process and by chiral  
symmetry.  The constraints of CS mandate that more complicated mechanisms
necessarily have larger values of $\nu$. Thus,  
provided that $E$ is smaller than some intrinsic energy scale,  
$\Lambda$, one has a decreasing series (i.e., it  
is a series in $E/\Lambda$) that is calculable.  
This series is organized in terms of  
$\nu$ rather than around unspecified (large) coupling constants.   
For this to work effectively $\Lambda$ must be sufficiently large,  
and more complicated mechanisms must not produce smaller  
values of $\nu$.  That is, amplitudes from loops and other products of
higher-order perturbation theory (PT) should not be larger than lower orders. 
As noted by Weinberg, the derivative structure of his
Lagrangian (which enforced CS) guarantees this.  Concomitant  
with the nonrenormalizability was the appearance of more and  
more unknown constants at higher orders in PT, which must be determined 
from experiment or calculated from QCD.  This feature clearly  
is a drawback, and one hopes that meaningful results can be  
obtained before the number of terms becomes too large for  
tractability.

This seminal, but skeletal, argument was developed into a highly  
successful program by the Bern group (and others)\cite{17} who  
introduced fermions.  A very important addition to the counting  
argument was made by Manohar and Georgi\cite{7}, which we will  
discuss later.  Much later (in a series of papers) Weinberg\cite{11} applied  
the procedure to nuclei.  Different terse versions of the counting were  
constructed in the various papers, while adapting the  
argument to the nuclear environment, a nontrivial achievement.   
This was extended in several ways in the thesis of van Kolck\cite{12},
which is an excellent introduction to those aspects of the problem that
we will ignore.

Our task is to reprise the derivation of Weinberg, add expository  
material familiar to nuclear physicists in order to make the results  
concrete, and finally to extract a number of crisp conclusions with  
import for nuclear physics and especially for the few-nucleon  
field.  It can be fairly stated that there is nothing in this primer that  
is not stated or implied in the work of Weinberg, van Kolck, and  
others.  Nevertheless, the importance of these ideas and their  
unfamiliarity to our field make an exposition desirable.  We will  
see that many qualitative results follow from simple arguments,  
once the framework has been constructed.

In order to satisfy the whims of the author, an older alternative  
form of power counting (the ``rules of scale''\cite{18,19}) will also be  
separately presented. Although this has been in use for decades\cite{20}, it 
is inappropriate for such complex mechanisms as loops and, moreover, it has no  
grounding in chiral symmetry.  It is, however, rather simple  
minded, easy to use, and does generate some insight into the  
nuclear physics that is less visible in the more formal approach.

Finally, we will fashion all of our arguments around Feynman  
diagrams, so we assume that the reader has some familiarity with  
their structure\cite{21}.  Specific examples with regularization and  
renormalization are relegated to Appendices B and C.  The first example 
illustrates a number of currently popular techniques and approaches,
while the second treats a simple two-nucleon problem. An exposition of  
nuclear matrix elements in momentum space is extremely helpful  
in interpreting the power counting in nuclei, and is relegated to  
Appendix A.  Otherwise, we will attempt to motivate the physics  
as we proceed.

We will show the following results at various points in the primer:\\

\indent $\star$ The kinetic energy/nucleon scales as $Q^2/\Lambda$, where $Q$
is an effective momentum in the nucleus.\\
\indent $\star$ The ``intrinsic'' potential energy/pair scales as -$Q^3/f_\pi 
\Lambda$, where $f_\pi$ is the pion-decay constant.\\
\indent $\star$ The cancellation of these two comparable energies allows 
nuclei to be weakly-bound systems ($\langle T \rangle \sim - \langle V 
\rangle$ and binding energy $\ll$ mass).\\
\indent $\star$ Large strong-interaction coupling constants are caused by the
mismatch between the $f_\pi$ and $\Lambda$ scales.\\
\indent $\star$ $N$-nucleon forces decrease in strength as $N$ increases. This 
powerful result authenticates decades of nuclear physics supposition and 
phenomenology.\\
\indent $\star$ Increasingly more complicated forces contribute more 
weakly.\\
\indent $\star$ Arbitrary processes contribute to the energy as $Q^\nu$, 
where $\nu = 1 + 2 (n_c + L) + \Delta$, with $L$ the number of loops, 
$n_c$ a topological parameter, and $\Delta \geq 0$ reflects the complexity 
of the interaction. This is Weinberg power counting.\\

\section{\underline{Rules of Scale - Nuclear Methodology}}

A type of primitive power counting has been in use for two decades\cite{18,19,
20}. The ``rules of scale'' were developed as a way to control expansions  
that arise in nonrelativistic treatments of meson-exchange  
currents.  Specifically, expansions in powers of $1/c^2$ were organized by 
noting that the velocity of a typical nucleon (with mass $M_N$ and momentum 
$p$) is $v = p/M_N$, and consequently $v/c \sim p/M_N \, c$. Thus, counting  
powers of $1/M_N$ in an expansion is equivalent to counting powers of
$1/c$.  It was noted that nuclei are weakly bound, implying that  
the kinetic and potential energies satisfy $\langle T \rangle
\sim - \langle V \rangle$.   
Hence, $V$ should be counted as $1/M_N$, since $V \sim T \sim p^2/2 M_N$.
This scheme works well at ``tree'' level (i.e., those mechanisms that don't  
involve loops, which were typically of interest in those times) and,
more importantly, it works for the current continuity equation,  
which is essential for EM processes.  That is, consistency in that equation
demands that energies of all types be treated on the same  
footing.  Conversely, the scheme has no predictive power for  
loops.

In order to progress further, we need some indication that $v/c \ll 1$.   
Otherwise, the Schr\"{o}dinger approach is meaningless.  We can  
use the uncertainty principle and the radii\cite{22} of the few-nucleon  
systems ($\sim$ 1.5 - 2.0 fm) to produce:  $p c \sim \hbar c/R \sim$ 100-150  
MeV $\sim m_{\pi} c^2$, where the last relation is a convenient  
{\it mnemonic}, rather than a statement of principle, since $m_\pi$ (unlike 
$f_\pi$, for example) vanishes in the chiral limit.  This leads to
$$
v/c \sim p/M_N c \sim m_{\pi}/M_N \sim 10-20\%\, , \eqno (3.1)
$$
and $(v/c)^2$ is satisfyingly small on average.  In order to  
conform to a more standard notation, we will use $Q \, (\sim m_{\pi})$
rather than $p$.  We will see later that in processes involving  
pions this is indeed a typical scale\cite{23}.

At this point a little history is instructive.  There are at least three  
ways of organizing the interactions of pions with nucleons.  This  
familiar argument has been distracting and ultimately unproductive 
(the author has been guilty of this\cite{19,20,24}), but it illustrates a 
number of useful points.  Parity and time-reversal-invariance arguments allow 
the Lagrangian for a single pion-nucleon interaction to be either
$$
{\cal{L}}_{\pi NN} = - i G \bar{N} \gamma_5 \mbox{\boldmath $\tau  
\cdot \pi$} N\, , \eqno (3.2)
$$
or
$$
{\cal{L}}^{\prime}_{\pi NN} = - \frac{f}{m_{\pi}} \bar{N} \gamma_5  
\slash{\partial} (\mbox{\boldmath $\tau \cdot \pi$}) N\, , \eqno(3.3)
$$
where $f = G m_{\pi} / 2 M_N$.  These two forms are usually  
referred to as PS (pseudoscalar) and PV (pseudovector).  An  
alternative form is introduced by using the Goldberger-Treiman\cite{25}
relation:
$$
\frac{G}{M_N} = \frac{g_A}{f_{\pi}}\, , \eqno (3.4)
$$
where $G \simeq 13, g_A \simeq 1.26$ is the axial-vector coupling  
constant and $f_{\pi} \simeq 92.4$ MeV is the pion-decay constant.  This  
remarkable relationship (the author's favorite in strong-interaction  
physics) relates strong interactions on the left side to weak  
interactions on the right, and is violated at the level of 2\% \cite{26} by
chiral-symmetry breaking (the left side is larger).  Thus we can  
also rewrite Eq. (3.3) by replacing $f/m_{\pi}$ with $g_A/2f_{\pi}$,
generating the third and preferred form
$$
{\cal{L}}^{\prime \prime}_{\pi NN} = - \frac{g_A}{f_{\pi}} \bar{N} \gamma_5  
\slash{\partial} (\mbox{\boldmath $t \cdot \pi$}) N\, , \eqno(3.5)
$$
where we have written $\mbox{\boldmath $t$}$ = $\mbox{\boldmath $\tau$}$/2
(a very common practice).

Note the very different sizes of the interactions in Eqs.\ (3.2) and
(3.3):  $G \sim 13$, while $f \sim 1$.  These two forms are  
equivalent on-shell, but differ dramatically off-shell.  We will see  
later that the off-shell behavior is controlled by chiral symmetry  
through terms we have not explicitly written.  In particular the  
large $N\bar{N}$-``pair'' terms implicit in the first form are exactly  
cancelled, a phenomenological result known historically as ``pair  
suppression''.  These pair terms have a very large dimensionless strength,
$G^2 m_{\pi} /  M_N \sim 25$, producing many-body forces that would make
the tractability of nuclear physics problematic.
The large size of $G$ is typical of the strong interactions, while  
$f$ is much more modest.  Even more modest is the effective  
coupling constant for the one-pion-exchange potential (OPEP)  
shown in Fig.\ (1a).  This amplitude, when converted to  
configuration space, generates a factor of $1/4 \pi$, whose  
precise form depends on the fact that we live in three space  
dimensions (see Appendix B):
$$
\int \frac{d^3 q}{(2 \pi)^3} \frac{e^{i\mbox{\boldmath  
$q \cdot r$}}}{\mbox{\boldmath $q$}^2 + m^2_{\pi}} = \frac{m_{\pi}}{4  
\pi} \left[ \frac{e^{-m_{\pi}r}}{m_{\pi}r} \right]\, . \eqno(3.6)
$$
\begin{figure}[htb]
\epsfig{file=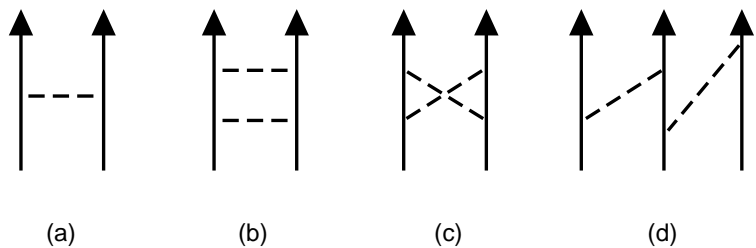,height=1.5in}
\caption{One- and two-pion-exchange contributions to the nuclear force.
Solid lines are nucleons, while dashed lines are pions.}
\end{figure}

Rendering OPEP\cite{9} into a dimensionless radial function of ${\bf x} = 
m_{\pi} {\bf r}$ and an overall dimensionful factor, we have $V_{\pi}  
(\mbox{\boldmath $r$}) \equiv \frac{m_{\pi}^3 g_A^2}{4 \pi f_\pi^2} V(\bf{x})$. 
Obviously some care is required in this exercise, since $4 \pi$ is  
dimensionless.  It is, however, a large number and we extract it.   
We expect $\langle V ({\bf x}) \rangle \sim 1$ since $ \langle 1 \rangle 
\equiv 1$, where $1$ is the unit operator (this is our normalization 
convention) and because
correlations with a length $\sim 1 / m_{\pi}$ are introduced into  
the wave function by $V (x)$. This implies that values of x near 1 are very 
important (which can be verified in Fig.\ (7)) and our functions near $x=1$ are
${\cal O}$(1) [cf., Eq.\ (3.6)]. The net result is that we expect a  
pair of nucleons exchanging a pion to generate an energy $\sim  
m_{\pi}^3 g_A^2 / 4 \pi f_\pi^2 \sim$ 20 MeV.  Anticipating a 
result from Section 5 and Appendix B, we note that $4 \pi f_{\pi} \sim$ 
1200 MeV $\sim \Lambda$, where $\Lambda$ was introduced in Section 1.
Recalling $Q \sim  
m_{\pi}$, we obtain $\langle OPEP \rangle \sim \frac{Q^3}{ \Lambda f_{\pi}}C_V$ 
per pair, where $C_V$ is the product of all the dimensionless factors $\sim 1$
(by supposition). We have already argued that $\langle T \rangle \sim  
\frac{Q^2 C_T}{\Lambda}$ per nucleon, where $C_T \sim 1$ and we have used
$M_N \sim \Lambda$.
For the triton, $\langle V_{\pi}\rangle \sim -15$ MeV/pair and $\langle T\rangle
\sim 15$ MeV/nucleon\cite{23}.  For the $\alpha$ particle both energies are  
somewhat larger.  In heavier systems, where the Pauli Principle  
begins to play a large role, these numbers (particularly the  
potential energy) cannot be correct.  Nevertheless, the ``intrinsic''  
scale for these quantities is
$$
\langle T \rangle / {\rm nucleon} \sim Q^2 / \Lambda\, , \eqno (3.7)
$$
$$
\langle V_{\pi} \rangle / {\rm pair} \sim \frac{-m^3_\pi}{(4 \pi f_\pi) f_\pi} 
\sim \frac{- Q^3}{\Lambda f_{\pi}} \sim \frac{- Q^2}{\Lambda}\, , \eqno (3.8)
$$
confirming that kinetic and potential energies (and quantities derived 
from them, such as impulse-approximation and meson-exchange EM currents)
should be counted similarly (numerically, $Q \sim f_{\pi}$), in the  
absence of large dimensionless coefficients that would skew the similarity.
Thus, for OPEP and light nuclei we see a confirmation of the weak-binding 
hypothesis that formed the basis for the rules of scale. Moreover, the
numerical results based on calculations with realistic potentials are 
consistent with these scales.

Two other potentials have been calculated that can be similarly  
analyzed.  Figures (1b) and (1c) show various parts of the
two-pion-exchange potential (TPEP).  These Feynman graphs  
can be separated into two components:  an intrinsic TPEP and  
the simple iteration of OPEP (which happens automatically  
whenever the Schr\"{o}dinger equation is solved).  The ``intrinsic''  
size of the former is\cite{24}
$$
\langle TPEP \rangle \sim \frac{m_{\pi}^5}{(4 \pi f_\pi)^3 f_\pi} \sim  
\frac{Q^5}{\Lambda^3 f_{\pi}} \sim \frac{1}{2} \, {\rm MeV / pair}\, . 
\eqno (3.9)
$$
Two-pion-exchange three-nucleon forces ($2 \pi E 3 NF$)  
can be similarly analyzed\cite{19} (see Fig.\ (1d)):
$$
\langle 2 \pi E 3 NF \rangle \sim \frac{m^6_{\pi}}{M_N (4 \pi f_\pi)^2 f^2_\pi} 
\sim \frac{Q^6}{\Lambda^3 f^2_{\pi}} \sim \frac{3}{4}\, {\rm MeV / triplet}\, .
\eqno (3.10)
$$
In counting arguments of this type, we have avoided factors of  
$g_A, 2, 1/2$, etc.,  and hope that they average out $\sim$ 1.  In most
cases, they do (within an overall factor of 2).  In Sections 6, 8, and 9, 
we will derive a very simple formula (Weinberg's formula) that  
reproduces all of these results by power counting (in $Q$ and $1 /  
\Lambda$) with almost no effort.

The only remaining problem is Fig.\ (1b).  This diagram, unlike a typical
case, does not have momenta $\sim Q \sim m_{\pi}$ flowing  
through every propagator.  Indeed, after the first interaction the  
propagator is typically $(E - T)^{-1} \sim (p^2 /2 M_N)^{-1} \sim M_N/Q^2$.  
Graphs of this type are called ``reducible''.
The small-$Q$ or ``infrared'' singularity enhances the graph by a factor 
$\sim M_N/ m_{\pi}$ compared to a typical case.  This singularity is also 
the origin of the ``ambiguity problem''\cite{19,20,24},
which reflects the fact that a potential is an off-shell amplitude, and
hence is not unique. The dichotomy between these two types of graphs
(reducible and irreducible) greatly reinforces the case for the following 
calculational scenario:

\indent $\diamond$ Calculate irreducible graphs (where simple  
power counting works) and define this as the nuclear potential.

\indent $\diamond$ Solve the Schr\"{o}dinger equation with that
potential -- the infrared enhancements will happen automatically.   

On the basis of several examples (this does not in any sense  
constitute a proof, which will follow in Section 6) we find that\\

\indent $\bullet$ Various nuclear energies behave as $Q^{\nu} /  
\Lambda^{\nu - n_c - 1} f^{n_c}_{\pi}$ for some values of $\nu$ and $n_c$, with 
more complicated mechanisms generating larger values of $\nu$. We have used 
$\Lambda \sim 4 \pi f_\pi \sim M_N$ in this simple counting exercise.

\section{\underline{Effective Interactions}}

The basic idea is a simple one.  Strong-interaction physics can be  
divided into two parts:  short-range (high-energy) parts and  
long-range (low-energy) parts.  We hope that the  
former is not dominant, but it could be.  This is the most difficult  
domain of nuclear physics, where phenomenology reigns  
supreme.  The long-range physics is of two (nonexclusive) types:  (1)  from  
pion-range physics, and (2) from iterations of the nuclear potential, as 
we saw in the previous section. Even though convergence requirements 
restrict us to the low-energy regime, it is essential that the pion  
degrees of freedom be treated explicitly; they have a strong energy 
dependence even at low energy.  We also saw that the infrared singularity 
that results
from iterating the potential plays a very important role in nuclear physics.

\begin{figure}[htb]
\epsfig{file=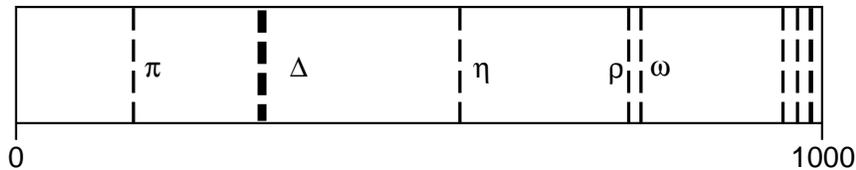,height=1.1in}
\caption{Unflavored meson spectrum [plus $\Delta$-isobar excitation energy] 
below 1 GeV.}
\end{figure}

The short-range regime is where all the complexity of QCD  
resides.  Consider, for example, the unflavored meson spectrum  
below 1 GeV, plotted in Fig.\ (2). Only two mesons lie below 770 MeV:  
$\pi$ and $\eta$.  In an SU(3) (rather than SU(2)) treatment of the chiral 
symmetry (i.e., strangeness is included), we would treat the ($\pi,  
\eta, K$)-meson octet together.  Although the $\eta$-meson  
contribution to the nuclear force is not particularly important, in an  
SU(2) approach its low mass would be problematic for the formalism.   
The other (heavy) mesons have masses $\geq$ 770 MeV.  We can therefore
imagine a scheme where the effect in appropriate spin and isospin 
channels of {\it all} single and multiple heavy-meson exchanges, loops, 
$\ldots$ is ``frozen out''\cite{15}.  That is, the effects of all possible  
short-range mechanisms are lumped together without regard to
their origin.  This is accomplished by noting that (for low energies) a  
derivative expansion about the zero-range limit is formally possible.
Thus $\Lambda \sim m_\rho \sim$ 1 GeV is also the boundary between the
short-range and long-range physics.

\begin{figure}[htb]
\epsfig{file=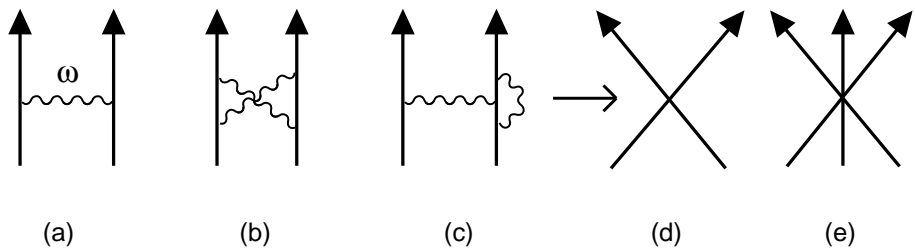,height=1.5in}
\caption{$\omega$-meson contributions in (a-c) to short-range $NN$ force in
(d), plus an additional 3NF in (e). Solid lines are nucleons, while wavy 
lines are $\omega$ mesons.}
\end{figure}

Figure (3) illustrates this approach, where a few selected mechanisms
involving $\omega$ mesons are shown.  All are subsumed in (3d),
which is the sum of various zero-range interactions that can  
include derivatives.  As an example, we treat Fig.\ (3a), for which the 
S-matrix is given by
$$
S = -i g^2_{\omega} \frac{(\bar{u}_1 \gamma_{\mu} u_1)(\bar{u}_2  
\gamma^{\mu} u_2)}{q^2 - m^2_{\omega}}\, , \eqno (4.1)
$$
where we follow the conventions and metric of Ref.\ \cite{21}.  We have dropped
normalization factors and the overall four-momentum-conserving factor, 
$(2 \pi )^4 \delta^4 (P_1 +P_2
-P_1^\prime -P_2^\prime)$, for this two-nucleon cluster. The  
strong coupling of the $\omega$ (with mass $m_{\omega}$) to the  
nucleon is reflected by the coupling constant, $g_{\omega}$, the  
nucleons (with spinors, $u$) have been arbitrarily labeled ``1'' and  
``2'', and $q^{\mu}$ is the momentum transferred in the collision  
($q^2 < 0$).  We now expand the denominator for $-q^2 \ll m^2_{\omega}$
$$
\frac{1}{m^2_{\omega} - q^2} \simeq \frac{1}{m^2_{\omega}} +  
\frac{q^2}{m^4_{\omega}} + \cdots \, , \eqno (4.2)
$$
and define each term in the series as a new {\it effective}  
interaction (Lagrangian)
$$
S \simeq i \left( \frac{g^2_{\omega}}{m^2_{\omega}} \right)  
(\bar{u}_1 \gamma_{\mu} u_1) (\bar{u}_2 \gamma^{\mu} u_2) (1 +  
\frac{q^2}{m^2_{\omega}} + \cdots )\, , \eqno (4.3)
$$
leading to (see Ref.\ \cite{21x})
$$
{\cal{L}}_{eff} = {\cal{L}}^{(0)}_{eff} + {\cal{L}}^{(2)}_{eff} + \cdots\, ,
\eqno (4.4)
$$
$$
{\cal{L}}^{(0)}_{eff} = \frac{1}{2} \left(  
\frac{g^2_{\omega}}{m^2_{\omega}} \right) (\bar{N}  
\gamma_{\mu} N)^2 \, . \eqno (4.5)
$$

We have defined the effective Lagrangian so that the usual  
Feynman rules produce, term-by-term, the same S-matrix.  We have added 
a superscript to $\cal{L}$, $``\Delta" = 0, 2, 4, \ldots$, that serves  
the same purpose as Weinberg's $``\nu"$ that we introduced  
earlier.  It classifies the interaction here according to powers of $q$;
we will develop the rules  
later for this notation. More complicated short-range forces are possible,
as indicated by the three-nucleon force in Fig.\ (3e).

We have used the fact that the $\omega$ meson is sufficiently  
heavy that (for small $q^2$) it propagates only a very short distance between  
emission and reabsorption (\`{a} la the uncertainty principle)\cite{27}.
Since we have assumed a particular mechanism, we have  
developed a {\it model} for the strength of the zero-range  
interaction.  Unfortunately, a vast array of mechanisms including
but not restricted to Figs.\ (3a) - (3c) (viz., loops to arbitrary order, 
physics not involving the $\omega \ldots$) also contribute.  Thus, this 
coefficient (and others with $\Delta > 0$) is unknown and must be determined  
phenomenologically or calculated directly from QCD.

Two more aspects are important and somewhat controversial.  In  
addition to mesons being ``exchanged'', we also have to consider  
nucleon resonances.  Figure (2) shows the $\Delta$-isobar  
excitation energy superimposed on the meson spectrum.  The  
low excitation energy of the $\Delta$ has long stimulated the  
imagination of nuclear theorists, and mechanisms proposed for  
almost every nuclear effect will typically include a $\Delta$  
somewhere.  The uncertainty principle states that the virtual  
excitation of a $\Delta$ scales inversely with the $\Delta$ energy (i.e., 
as $1/E_{\Delta}$).  Recent work suggests that in some regimes 
(viz., three-nucleon forces and meson-exchange currents\cite{28}) $E_{\Delta}$ 
behaves more as $\Lambda \sim$ 1 GeV than $M_{\Delta} - M_N \sim$ 300 MeV.  In 
other regimes (threshold $pp \rightarrow pp \pi^0$, for example\cite{29}) 
the opposite is more likely.  In constructing effective interactions, some 
theorists\cite{12} routinely
include the $\Delta$ in the active Hilbert space, while others do  
not.  Care should be taken when assuming that the $\Delta$ is  
not very important.  We will, however, ignore the $\Delta$ in what  
follows solely for reasons of simplicity. 

The second aspect concerns relativity.  The simple example  
involving the $\omega$ meson respected special relativity,  
including the derivative terms ($\sim q^2$).  More complex  
mechanisms would lead to terms proportional to the overall  
momentum $p^{\mu}$ of a nucleon, rather than the momentum  
transfer, $q^{\mu}$.  While the latter is typically not large,  
$p^{\mu} = (E, \mbox{\boldmath $p$})$, and $E \sim M_N \sim \Lambda$ is large.  
This would lead to a series of terms that
would all be the same size (recall that we have a series in $E/\Lambda$).
The problem can be formally eliminated by  
performing a $1/M_N$ expansion (e.g., a Foldy-Wouthuysen
reduction\cite{20,21}), where the effect of $N \bar{N}$ ``pair'' terms is  
projected out, and one treats only the positive-energy spectrum. This 
``freezing out'' of degrees of freedom is familiar to nuclear physicists in 
the Feshbach (P-Q) reaction theory\cite{31}, where the Hilbert space is 
compacted from the normal to a restricted size, leading to much more  
complex operators (which must reproduce the physics of the  
larger Hilbert space in the smaller one).  Covariance is lost, but one 
retains an  
expansion in $Q$, rather than introducing terms $\sim M_N$ at  
every order.  This alternative formalism is given the generic name  
``heavy-baryon'' and is currently the method of choice\cite{32,34}.  This  
presupposes that a nonrelativistic treatment of the nuclear  
physics is appropriate, which clearly holds for the few-nucleon  
systems. We give an example of a nonrelativistic loop calculation
in Appendix B, and compare it to a relativistic one. Appendix C treats a simple
nonrelativistic two-nucleon model.

We summarize this section by noting that:\\

\indent $\bullet$ The complexities of strong-interaction physics  
can be divided into long-range and short-range parts, with pion  
d.o.f.\ dominating the former and the latter condensed into  
zero-range interactions of unknown size.

\indent $\bullet$ Effective interactions are constructed by  
``freezing out'' the short-range degrees of freedom (mesons,  
resonances, $N \bar{N}$ pairs, $\ldots$), leading to structurally more  
complex interactions that (hopefully) are easier to treat at low  
energies.

\indent $\bullet$ The $\Delta$ isobar can be included in the active  
Hilbert space (with the pions) or not, depending on the problem to be solved.

\indent $\bullet$ A nonrelativistic, nonrenormalizable field theory  
is possible to construct and to use.

\section{\underline{Dimensional Power Counting in Lagrangians}} 

We begin by considering the dimensions of various quantities in order to assess
the scales of the strong interactions. Our approach in this section will be 
to motivate rather than to attempt rigor. The interested reader should consult 
Refs.\ \cite{1,7,8} for a more sophisticated approach. We will also formulate 
the arguments supposing a relativistic field theory. Following the discussion 
of the previous section, they also apply {\it mutatis mutandis} to a 
nonrelativistic ``heavy-baryon'' field theory.

We denote a power $\nu$ of energy by $[E^\nu ]$ (equivalent to $length^{-\nu}$),
and recall that a Lagrangian
has dimension 4 (i.e., behaves as $[E^4]$).
The Lagrangian for a system of pions and nucleons interacting  
through derivatives ($\partial^{\mu}$) acting on any field can be  
written in the schematic form
$$
{\cal{L}} = {\cal{L}}_{free} + \Delta {\cal{L}}\, , \eqno (5.1)
$$
$$
{\cal{L}}_{free} = \bar{N} (i \gamma_\mu \partial^\mu - M_N) N + \frac{1}{2}  
(\partial^\mu \mbox{\boldmath $\pi$})^2 - \frac{1}{2} m^2_{\pi}  
\mbox{\boldmath $\pi$}^2\, , \eqno (5.2)
$$
$$
\Delta {\cal{L}} \sim a \left( \frac{\bar{N} (\cdots) N}{b} \right)^{\ell}
\left( \frac{\mbox{\boldmath $\pi$}}{c} \right)^m 
\left( \frac{\partial^\mu}{d} \right)^n\, . \eqno (5.3)
$$
The combination $\bar{N} N$ has dimension $[E^3]$, while  
$\mbox{\boldmath $\pi$}$ and $\partial^{\mu}$ (as well as $m_\pi$ and $M_N$)
have dimension $[E]$, and $\Delta {\cal{L}}$ of course has dimension $[E^4]$. 

Our first assumption is that only the scales $f_{\pi}$ and $\Lambda$ 
occur in $a, b, c, d$ and also that $\Delta {\cal{L}}$ should reproduce 
${\cal{L}}_{free}$.  Clearly, $c$ must be $f_{\pi}$, since that quantity 
sets the scale for the pion field (cf., PCAC\cite{1,2,6,21}).  Setting 
$\ell = 0$ and $m = n = 2$ gives $a / c^2 d^2 \sim 1$, 
while $m = 0, \ell = 1$, and $n = 0, 1$ give $a / b \sim \Lambda$ and 
$a / bd \sim 1$ (where we have substituted $M_N  
\sim \Lambda$).  These three equations can be uniquely solved  
to produce $d \sim \Lambda, a \sim f^2_{\pi} \Lambda^2$, and $b 
\sim f^2_{\pi} \Lambda$ and thus
$$
\Delta {\cal{L}} = c_{\ell m n} \left(\frac{\bar{N} (\cdots) N}{f^2_{\pi}  
\Lambda} \right)^{\ell} \left(\frac{\mbox{\boldmath  
$\pi$}}{f_{\pi}}\right)^m \left( \frac{\partial^{\mu}, m_{\pi}}{\Lambda}  
\right)^n f^2_{\pi} \Lambda^2\, , \eqno (5.4)
$$
where $c_{\ell m n}$ is dimensionless, and we have used the  
form of ${\cal{L}}_{free}$ to incorporate the  
chiral-symmetry-breaking pion-mass term into Eq.\ (5.4): a pion mass 
(either explicit or implicit) is treated as a derivative.  All 
Dirac matrices, nucleon isospin  
operators, etc.\ have been ignored and are indicated by the dots.

This lovely formula has profound implications if we invoke
chiral symmetry\cite{8,35}. No unique representation exists for 
incorporating chiral symmetry (cf., PS vs. PV forms), but  a representation 
always exists in terms of covariant derivatives\cite{1,2,8,12},
which has only increasing powers of ($1 / \Lambda$).   
According to Eq.\ (5.4), these powers have the form $(1 / \Lambda)^{\Delta}$,
where
$$
\Delta = n + \ell - 2\, . \eqno (5.5)
$$
The minimum value for pions is $0$ from Eq.\ (5.2) and for nucleons 
is $-1$ for each of the two (largely) cancelling terms in Eq.\ (5.2) for  
free nucleons (i.e., $\gamma_0 E - M_N$).  This cancellation highlights 
the problem with derivatives of nucleon  
fields that we raised earlier: a slowly-moving nucleon actually  
has a kinetic energy, ${\cal{L}}_{free} = -N^{\dagger} 
\frac{\mbox{\boldmath $p$}^2}{2 M_N} N$, or $\Delta = 1$.  
The PV (derivative-representation) Lagrangian has $\Delta = 0$. These 
examples motivate the rigorous result\cite{8} that the 
derivative-representation chiral constraint has a form guaranteeing 
only positive powers of $1/\Lambda$ in Eq.\ (5.4):
$$
\Delta \geq 0\, . \eqno (5.6)
$$
The PS form in Eq.\ (3.2) has $\Delta = -1$, which tells us that  
additional terms in the Lagrangian are required in order to construct a  
proper representation of chiral symmetry.  Indeed, it was the  
neglect of the latter terms that led to the {\it ad hoc} ``pair suppression'' 
mechanism many years ago. This mechanism is automatic for chiral models or  
theories, as noted in many places.  The use of a chiral representation 
corresponding to Eq.\ (3.2) requires an additional (leading-order) 
``pair''-killing interaction of the form\cite{19}:
$$
\Delta {\cal{L}}_{PS} = \frac{G^2}{2M_N} \bar{N} \mbox{\boldmath  
$\pi$}^2 N\, . \eqno (5.7)
$$
We strongly argue that derivative forms be used, because the chiral constraint,
Eq. (5.6), applies term-by-term.

The second assumption is that a reasonable theory should have\cite{36}
$$
c_{\ell mn} \sim 1\, , \eqno (5.8)
$$
or that the theory is ``natural''. This is also called naive dimensional 
power counting (NDPC)\cite{7,35}.   Clearly, if the scaling in Eq. (5.4) is a  
figment of our imaginations, the values of $c_{\ell m n}$ will jump  
all over.  Verifying naturalness validates Eq.\ (5.4).  We also note  
that if a symmetry exists, $c_{\ell m n}$ could be vanishingly small, but
if the scaling hypothesis holds, very small coefficients would otherwise be 
just as unlikely as very large ones. 

The third assumption is that vacuum fluctuations do not alter the scales that 
we introduced. It is shown in Refs.\ \cite{1,7} that it is necessary for loop
integrals to be cut off at energies
$$
\Lambda \lsim 4 \pi f_{\pi} \eqno (5.9)
$$
in order for the structure in Eq.\ (5.4) to be preserved, and we  
indeed have used $4 \pi f_{\pi} \sim \Lambda$ interchangeably in  
Section 3 (see also Appendix B).

At this point, examples will serve us best in assessing the concept of  
``natural'', and to illustrate the use of Eq.\ (5.4). Evaluating  
that equation for $\ell = m = n = 1 (\Delta = 0)$ and comparing it to 
Eq.\ (3.5) produces
$$
c_{\pi} \sim g_A \sim 1.26 \, . \eqno (5.10)
$$
Using $\mbox{\boldmath $\tau$}/2$ instead of $\mbox{\boldmath $t$}$
would have produced the equally satisfactory $g_A /2$.

The zero-range $NN$ interaction  
produced by an $\omega$-meson exchange was derived earlier (Eq.\ (4.5))  
and can be compared to the case $\ell = 2, m = n = 0$, yielding
$c_\omega/f^2_\pi$ and a natural coupling constant
$$
c_{\omega} \sim f^2_{\pi} \frac{g^2_{\omega}}{2\, m^2_{\omega}}  
\sim 1.75\, , \eqno (5.11)
$$
using the numerical entries in Table (A.2) of Ref.\ \cite{37}.  A famous 
example is the KSFR relation\cite{38}, which states
$$
\frac{g_{\rho}}{m_{\rho}} \simeq \frac{1}{\sqrt{2} f_{\pi}}\, , \eqno (5.12)
$$
and corresponds to a zero-range Lagrangian:  $\frac{g^2_{\rho}}{2  
m^2_{\rho}} (\bar{N} \gamma_{\mu} \mbox{\boldmath $t$} N)^2 =  
\frac{1}{4 f^2_{\pi}} (\bar{N} \gamma_{\mu} \mbox{\boldmath $t$}  
N)^2$ or $c_{\rho} \sim 0.25$.  In general a  
heavy-meson exchange with mass $m_x$ and coupling constant  
(to the nucleon) $g_x$ corresponds to a zero-range Lagrangian with a  
coefficient, $\frac{g^2_x}{2 m^2_x} \sim \frac{c_x}{f^2_{\pi}}$,  
implying that
$$
g_x \sim \frac{m_x}{f_{\pi}} \sqrt{2 c_x} \sim \frac{\Lambda}{f_{\pi}}  
\sim 10\, , \eqno (5.13)
$$
using $m_x \sim \Lambda$, and $\sqrt{2 c_x} \sim 1$.  This  
illustrates why strong-interaction coupling constants are large  
dimensionless numbers ($\sim 10$): the scales $\Lambda$ and $f_\pi$
are very different.

The next example of this type is a caution.  What if there were a  
mechanism such as an $NN$ force component of normal size 
mediated by the exchange of a light meson $(m \ll \Lambda)$?
This would correspond to a large value of $c_x$, and would likely  
lead in perturbation theory to a growing series whose structure  
violates Eq.\ (5.4).  This is analogous to the intruder-state  
problem in nuclear-structure theory\cite{38x}, where a low-energy state  
badly affects convergence of perturbation theory calculations.   
Another way of saying the same thing is that there would be  
another scale in the problem, which would violate the  
assumptions that led to Eq.\ (5.4).

We can also extend the NDPC to other forces. One isospin violation (IV)
mechanism\cite{12} is parameterized by the up-down quark-mass difference 
$m_d - m_u \equiv \epsilon\, (m_d + m_u)$, where 
$\epsilon = (m_d - m_u)/(m_d + m_u) \sim 0.3$. The factor of $m_d + m_u$ is 
proportional to $m^2_\pi$\cite{8} and
Eq.\ (5.4) can still be used, provided that we identify\cite{26}
$$
c_{IV} \sim \epsilon \, c\, , \eqno (5.14)
$$
where $c$ is ${\cal O}$(1) and $n \geq 2$.  That is, Eq.\ (5.4) has the 
$c_{\ell m n}$ replaced by $\epsilon\, c_{\ell m n}$, and there are at least
two {\it implicit} powers of $m_\pi$. Similarly, 
parity-violating forces have $c$ replaced by\cite{38xx}
$$
c_{PV} \sim \frac{G_F\, f^2_\pi}{\sqrt{2}}\, c\, , \eqno (5.15)
$$
where $G_F$ is the Fermi constant and $c$ is ${\cal O}$(1).

Another application is the nuclear density, which for nuclear matter has the 
``empirical'' value\cite{39} $\rho_{\rm NM} = 0.153 \,{\rm fm}^{-3} \simeq 1.5
f^3_{\pi}$.  The generic Lagrangian contains powers of $\left( \frac{\bar{N} 
(\cdots ) N}{f^2_{\pi} \Lambda} \right)$, the numerator of which 
is essentially the nuclear density $(N^\dagger N)$.  At nuclear-matter 
density, the Lagrangian series (in $\ell$) is therefore geometric in 
$\frac{1.5 f^3_{\pi}}{f^2_{\pi} \Lambda} = 1.5  
f_{\pi} / \Lambda \sim 1/7$, which is comfortingly small, so that  
the expansion presumably works at normal nuclear density.

Our final example takes us outside the realm of few-nucleon systems. One might 
ask whether there have been any calculations performed using a Lagrangian of 
zero-range form. Although this has not been done in few-nucleon systems, there
does exist one comprehensive Dirac-Hartree calculation\cite{40} for a set of 
57 nuclei using such forms (but without pion d.o.f). The various terms in 
the Lagrangian can be represented schematically by $\alpha (\bar{N}N)^2 +
\beta (\bar{N}N)^3 + \gamma (\bar{N}N)^4 + \delta (\partial^\mu \bar{N}N)^2$,
where various isospin and Dirac matrices sit between $\bar{N}$ and $N$. A total
of nine coupling constants were adjusted to fit the energies and radii of three
representative nuclei. The results are among the best ever obtained for 
such a comprehensive 
set of nuclei. If one uses Eq. (5.4) with $m=0$ and various values of $\ell$ 
and $n$, and compares to the expressions in Ref.\ \cite{40,41}, one obtains 
the results in Table (5.1). 

\begin{table}[htb]
\centering
\caption{ Optimized coupling constants for the Lagrangian of Refs.\ 
\protect\cite{40,41} and corresponding dimensionless coefficients and 
chiral expansion order.}
\vspace{18pt}
\begin{tabular}{|c|rc|c|c|}
\hline
Constant & Magnitude & Dimension & ${  }c_{lmn}{  }$ & Order \\ \hline \hline
$\alpha_{S}$ & -4.508${\times}10^{-4}$ & MeV$^{-2}$ & -1.93 & $\Lambda^{0}$ \\
$\alpha_{TS}$ & 7.403${\times}10^{-7}$ & MeV$^{-2}$ & 0.013 & $\Lambda^{0}$ \\
$\alpha_{V}$ & 3.427${\times}10^{-4}$ & MeV$^{-2}$  & 1.47  & $\Lambda^{0}$ \\
$\alpha_{TV}$ & 3.257${\times}10^{-5}$ & MeV$^{-2}$ & 0.56  & $\Lambda^{0}$ \\
$\beta_{S}$ & 1.110${\times}10^{-11}$ & MeV$^{-5}$  & 0.27  & $\Lambda^{-1}$ \\
$\gamma_{S}$ & 5.735${\times}10^{-17}$ & MeV$^{-8}$ & 8.98  & $\Lambda^{-2}$ \\
$\gamma_{V}$ & -4.389${\times}10^{-17}$ & MeV$^{-8}$ & -6.87 & $\Lambda^{-2}$\\
$\delta_{S}$ & -4.239${\times}10^{-10}$ & MeV$^{-4}$ & -1.81 & $\Lambda^{-2}$\\
$\delta_{V}$ & -1.144${\times}10^{-10}$ & MeV$^{-4}$ & -0.49 & $\Lambda^{-2}$\\
\hline
\end{tabular}
\end{table}

The unscaled coupling constants (whose subscripts refer to Dirac and isospin 
matrices) span 13 orders of magnitude. Most of the scaled $c$'s that result 
are numbers near one. The average of the $\gamma$ terms is also natural. 
The uncomfortably large difference is possibly due to a lack of sensitivity 
in that quantity to the data. If the tiny value of $\alpha_{TS}$
is not an artifact of the fitting process or of the neglect of pion-range 
physics (unknown at this time) 
it presupposes a symmetry of some kind. Nevertheless, scaling as predicted by
Eq.\ (5.4) is obvious. Improvements in the quality of the many-body techniques 
are not expected to alter this conclusion\cite{41}, but will alter each number.

We summarize this section by noting that:\\

$\bullet$ A two-scale hypothesis ($f_\pi$ and $\Lambda$) for the dimensional 
factors in the effective Lagrangian (plus chiral symmetry) suggests a 
convergent expansion of the Lagrangian series for normal nuclear conditions.

$\bullet$ This expansion has been organized so that vacuum fluctuations do not 
alter the form.

$\bullet$ The dimensionless coefficients in the Lagrangian are ``natural''
if they are ${\cal O}$(1), which we illustrated by several examples.

$\bullet$ Chiral symmetry guarantees that the large scale ($\Lambda \sim 4 
\pi f_\pi \sim m_\rho \ldots$) does not occur with positive powers.

\section{\underline{Power Counting in Amplitudes}} 

Power counting in amplitudes is a straightforward exercise, but somewhat  
tedious.  It also is dependent on the environment and on what  
one chooses to emphasize.  The result is worth the effort,  
however, since most of our previous results are subsumed by the  
very simple final forms.  We work with nuclei, bound states of  
$A$ nucleons.  Bound states are forever interacting, and one  
cannot simply separate interaction diagrams into connected and
disconnected parts and discard the latter.  Eventually each nucleon  
interacts and shares energy and momentum with the others.  In  
normal (textbook) applications disconnected diagrams don't  
contribute.  We emphasize that the sharing of momentum  
dominates the systematics in a nucleus.  The momentum that one nucleon  
takes from the ``bank'' is not available to the others.

The basic idea is to use our knowledge of the dimensionality of  
propagators, phase space factors, vertices, and delta functions to  
find the dimensionality of an amplitude, after eliminating all the  
coupling constants.  
Given this and knowledge of how the other scales (viz., $f_{\pi}$
and $\Lambda$) come into the problem, the coupling constants can be
put back and a complete scheme can  
be constructed.  We note that finding a momentum behavior $\sim Q^{\nu}$ does 
not necessarily imply a final $(Q/\Lambda)^\nu$ behavior.  We also choose to  
work in configuration space, and this choice means that, while counting
powers of momenta, the effect  
of {\it additional} phase space factors (besides those in loops)
needed to convert to that space must also be incorporated, and this has been
done in the derivation leading to Eq.\ (6.12). We will give examples 
later that spell  
out the differences, and Appendix A documents the various  
factors that normally arise.  Finally, we postpone dealing with  
reducible (infrared-singular) diagrams until later.  We first deal  
with the amplitudes, and then we will worry about the
coupling constants.

Our first concern is what we should calculate and what rules apply. In
general\cite{20}, the energy shift, $\Delta E$, in an interacting system is 
given in terms of the S-matrix (see Appendix B) by $S_{fi} \equiv \delta_{fi}
-i\, \Delta E \; [(2 \pi)^4 \delta^4 ( P_f - P_i)]$. The $\delta_{fi}$
factor for $N$ noninteracting particles is the product of 
three-momentum-conserving $\delta$-functions of each particle (see Eq.\ (6.35)
of Ref.\ \cite{21}) and behaves as $[E^{-3N}]$, while the energy shift (which 
is also the T-matrix) therefore behaves as $[E^{4-3N}]$. For a cluster of two 
nucleons ($N = 2$) this has the dimensions $[E^{-2}]$, in agreement with
Eq.\ (4.1) (the spinors and coupling constant in that equation are 
dimensionless). That form, as argued above, lacks the phase-space factor
$\sim [E^3]$ needed to convert to configuration space. Together these factors
produce $\sim [E]$, the correct dimension for an energy. We therefore use the 
same rules for an energy shift as for an S-matrix, stripping off the 
four-momentum-conserving $\delta$-function (if there is a single cluster) to 
produce $\Delta E$. Note that if there are $C$ separate clusters, there will 
be ($C-1$) cluster $\delta$-functions remaining to treat.

Our second concern is that there are many
possible quantities that specify how momentum and energy flow 
in a given diagram.  The key is to decide which ones to keep and which  
to eliminate.  Typically one eliminates internal variables and  
keeps external ones.  This is not enough for specificity and one  
additional choice remains.  That choice will be made in such a  
way that the chiral constraint, Eq.\ (5.6), can be implemented by inspection.   
We also will calculate for fixed $A$, and eliminate at the end  
some extraneous factors that are dependent on our wave-function
normalization scheme and depend only on $A$. Finally, in keeping with
common practice, we work in $D$ space-time dimensions $(D \equiv 4)$. 
For simplicity we restrict ourselves to no external bosons (interacting with 
our nucleus) and only calculate the energy shift.

\begin{figure}[htb]
\epsfig{file=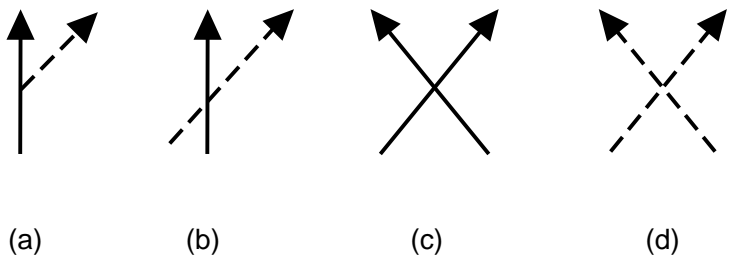,height=1.5in}
\caption{Elements of Lagrangian (i.e., vertices) for a system of pions and 
nucleons. Solid lines are nucleons, while dashed lines are pions.}
\end{figure}

The first set of variables specifies what happens at individual  
vertices or interaction points, and we assume that there is at least one
vertex.  Figure (4) contains several examples  
from previous sections plus an additional one.  At the $i^{th}$  
vertex in a Feynman diagram we have\\

$\diamond$ $b_i$ bosons (i.e., pions in our case) entering  
or leaving;

$\diamond$ $f_i$ fermions (i.e., nucleons) entering or leaving; \hfill (6.1)

$\diamond$ $d_i$ derivatives acting at the vertex;

$\diamond$ $ \Delta_i = d_i + f_i / 2 - 2\, \geq 0 \, , $\\

\noindent where the last definition anticipates our final form and corresponds 
exactly to Eq.\ (5.5), since $\ell = f_i / 2$ and $n = d_i$.   In Fig.\ (4a) 
we have $b = 1, f = 2$.  For the vertex corresponding to Eq.\  
(3.2), $d = 0$ and $\Delta = -1$, while $d = 1$ and $\Delta = 0$  
for Eq.\ (3.3).  Figure (4b) has $f = 2, b = 2$, while Eq.\ (5.7) has  
$d = 0$ and $\Delta = -1$, as we discussed before.  Figure (4d) is  
the Weinberg four-pion interaction\cite{8,11} and has 2 derivatives  
producing $f = 0, b = 4, d = 2$, and $\Delta = 0$.  Recall that the  
$\Delta = -1$ terms separately violate chiral symmetry and cancellations must 
occur between them. Figure (4c) and Eq.\ (4.5) have $f = 4$ and
$d = b = \Delta = 0$.

The second set of variables will completely specify the process:\\

\indent $\diamond$ $\nu =$ dimensionality of the amplitude without  
coupling constants $(Q^{\nu})$;

\indent $\diamond$ $D =$ number of space-time dimensions;

\indent $\diamond$ $A =$ number of nucleons (which is conserved);

\indent $\diamond$ $L =$ number of loops (i.e., internal phase-space  
integrals)\, ($\geq 0$);\hfill (6.2)

\indent $\diamond$ $n_c =$ number of nucleons interacting with  
{\it at least} one other minus the number of  
clusters with {\it at least} two interacting nucleons\, ($\geq 0$);

\indent $\diamond$ $\bar{L} \equiv n_c + L \, (\geq 0)$\, ;

\indent $\diamond$ $ \Delta \equiv \sum_i \Delta_i \,(\geq 0)\,  $
(i.e., sum over {\it all} vertices).\\

\begin{figure}[htb]
\epsfig{file=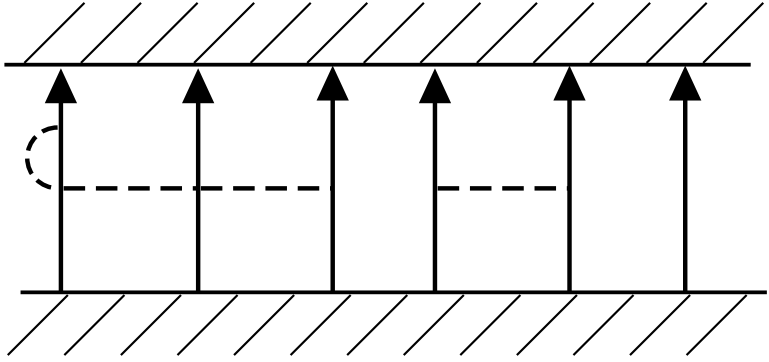,height=2.0in}
\caption{System of 6 nucleons interacting via pions. The cross-hatched 
(for emphasis) regions depict initial and final times. Solid lines are 
nucleons, while dashed lines are pions.}
\end{figure}

Figure (5) shows a representative case with the requisite  
complexity.  It has $A = 6$ nucleons propagating from some initial time to some 
final time (indicated by cross-hatching for emphasis). The process can be
divided into two interacting  
clusters and a single noninteracting nucleon, producing $n_c =  
3$.  There is a closed loop on the left-most nucleon, so that $L =  
1$ and $\bar{L} = 4$.  The very important topological parameter $n_c$  
specifies the complexity of the interaction scenario and will  
determine the relative importance of $N$-nucleon forces.

A third set of variables is required to set up the counting, but will  
not appear in the final result:\\

\indent $\diamond$ $E_N =$ number of interacting nucleons;

\indent $\diamond$ $X_N =$ number of noninteracting nucleons;

\indent $\diamond$ $C =$ number of interacting clusters of nucleons;\hfill (6.3)

\indent $\diamond$ $I_B =$ number of boson (pion) propagators (i.e., internal 
lines);

\indent $\diamond$ $I_F =$ number of fermion (nucleon) propagators (i.e., 
internal lines);

\indent $\diamond$ $V =$ number of vertices (i.e., $\sum_i$).\\

\noindent Note that a cluster must contain at least one vertex, but can have 
any number (including only one) of connected nucleons. Moreover,
the nucleon lines touching the cross-hatched regions are not propagators.
For the case specified by Fig.\ (5) we have $C = 2, E_N = 5, X_N =  
1, V = 6, I_B = 4$, and $I_F = 1$.  

We have several obvious relationships between previously defined variables
$$
A = X_N + E_N\, , \eqno (6.4)
$$
$$
n_c = E_N - C\, , \eqno (6.5)
$$
$$
V = \sum_i 1\, . \eqno (6.6)
$$
Note that Eq. (6.5) does not appear to correspond precisely to Eq. (6.2).
A cluster of one nucleon interacting only with itself will not contribute
to $n_c$, and the two definitions are thus equivalent.
Using the fact that two boson (pion) or fermion (nucleon) fields  
are needed to make each boson or fermion propagator, we have
$$
\sum_i b_i = 2I_B\, , \eqno (6.7)
$$
$$
\sum_i f_i = 2I_F + 2E_N\, , \eqno (6.8)
$$
where by definition only interacting nucleons connect to a vertex.   
The number of loops in a diagram can be counted by noting that  
in constructing a diagram there is a four-dimensional  
phase-space integral associated with each propagator.  Feynman  
diagrams conserve four-momentum at each vertex, and these  
constraints ($V$ of them) eliminate the integrals.  This  
overcounts, however, since each cluster has an overall $\delta$-function 
(four-momentum) constraint (see the discussion below Eq.\ (4.1)), and there 
are $C$ of them. Thus, the number of loops is
$$
L = I_B + I_F - V + C\, , \eqno (6.9)
$$
and associated with each loop is a four-(i.e., $D$-)dimensional phase-space 
integral $\int  
\frac{d^ D p}{(2 \pi)^D} \sim [E^D]$.  Figure (5) has one loop, by inspection 
or by using Eq.\ (6.9). Each noninteracting nucleon contributes a 
three-momentum  (i.e., $D-1$) $\delta$-function $\sim [E^{1-D}]$ 
(see Appendix A).

Finally, the momentum or energy dimensionality of any diagram  
can be determined by counting:\\
$\diamond$ phase-space factors $\sim [E^D]$ in loops;\\
$\diamond$ boson (pion) propagators ($\sim (p^2 - m^2_{\pi})^{-1} 
          \sim [E^{-2}]$);\\
$\diamond$ fermion propagators ($\sim (\slash{p} - M_N)^{-1} \sim [E^{-1}]$);\\
$\diamond$ derivatives in vertices ($p^{\mu} \sim [E]$);\\
$\diamond$ cluster $\delta$-functions $\sim [E^{- D}]$ ($C-1$ of them; one has 
already been removed) ;\\
$\diamond$ $\delta$-functions $\sim [E^{1 - D}]$ for noninteracting nucleons 
          (see above and App.\ A);\\
$\diamond$ an optional factor [in brackets below] (because it depends only on A,
          and not the process) that enforces the overall normalization of the 
          wave function in momentum space (see Appendix A).\\
In that order we have
$$
\nu = D L - 2 I_B - I_F + \sum_i d_i - D(C-1) - (D-1) X_N + [(D-1)(A-1)]\, .
\eqno (6.10)
$$

There are 5 basic internal ``variables'' in Eqs.\ (6.7 - 6.9) (together  
with $C, E_N, V$) that determine $\nu (I_B, I_F, L, b_i,  
f_i)$ and only those 3 relations among them; $I_B$ and $I_F$  
are always eliminated.  We must keep 2 of the remaining ones and choose 
$L$ and $f_i$.  The other choices produce equivalent,  
though less useful, formulae (see Ref.\ \cite{12} for a discussion of  
options).  Eliminating $I_B, I_F$ and $\sum_i b_i$ using Eqs.\  
(6.7 - 6.9) we find
\begin{eqnarray}
\nu &=& 1 + (D - 2) (n_c + L) + \sum_i \Delta_i - ((D - 1)(A-1)) 
+  [(D-1)(A-1)] \nonumber \\
&=& 1 + (D - 2) \bar{L} + \Delta \nonumber \, .\hspace*{2.75in} (6.11)
\end{eqnarray}
The constant term in square brackets cancels an
identical term that arises from algebraic manipulation. Note also that the 
last two terms in Eq. (6.10) in effect add 
($D-1$) (i.e., 3) powers (from phase-space factors) for each 
{\it independent} (i.e., non-center-of-mass) nuclear coordinate 
involved in an interaction. These powers
(as we will see below in an explicit example) in effect convert from 
momentum to configuration space.  Finally, for  
D = 4 we have a very simple and elegant power-counting result:
$$
\nu = 1 + 2 \bar{L} + \Delta\, . \eqno (6.12)
$$
Except for irrelevant constant terms (depending on $A$), this agrees with  
Weinberg and van Kolck\cite{11,12}.

Recall that $\bar{L}$ and $\Delta$ are separately positive semi-definite.
The case $\bar{L} = \Delta = 0\  (\nu = 1)$ cannot occur because this
would imply a single nucleon interacting without a loop; only the nucleon
kinetic energy has this form and it corresponds to $\Delta = 1$, as we
found in Eq. (3.7). Thus, the minimum values of $\nu$ are 2, corresponding
to the kinetic energy, and 3, corresponding to $NN$ forces ($n_c=1, L=0,
\Delta=0$). See also Appendix B for another example.

We summarize this section by noting that:\\

\indent $\bullet$ Powers, $\nu$, of a generic energy or  
momentum, $Q$, can be counted in Feynman diagrams by  
following the flow of momentum through each vertex.

\indent $\bullet$ The nuclear case requires consideration of {\it all}
nucleons, since all nucleons eventually interact in a bound state.

\indent $\bullet$ The final formula for $Q^{\nu}$ is exceptionally  
simple, and shows that $\nu$ {\it always} increases as more  
complicated mechanisms are considered.

\section{\underline{Electromagnetic Interactions}} 

We can extend these results to EM interactions within  
strongly-interacting systems by including the ``photon'' as an extra  
boson\cite{10}.  This produces only one significant change.  For EM  
interactions:  $\Delta^{EM}_i \geq -1$, as noted by Rho\cite{10}. This is  
compensated by a factor of $e$, the fundamental charge, which  
will reduce the size of amplitudes.  For specificity we will illustrate  
the case of a single photon (virtual or otherwise) and this will allow us to
discuss nuclear EM currents (either impulse-approximation or
meson-exchange), $J^{\mu}_{EM} \equiv (\rho,{\bf J})$. We refer the reader to  
Refs.\ \cite{10,26} for more complex cases.

Separating out the (assumed) single EM vertex from Eq.\ (6.12), we 
have\cite{10}
$$
\nu_{EM} = 1 + 2 \bar{L} + \Delta_{ST} + \Delta_{EM}\, , 
\eqno (7.1)
$$
where $\Delta_{ST}$ is the sum over strong vertices ($\geq 0$) and 
$\Delta_{EM}$ refers to the single EM vertex $(\geq - 1)$ and will differ for 
$\rho$ and ${\bf J}$.

\begin{figure}[htb]
\epsfig{file=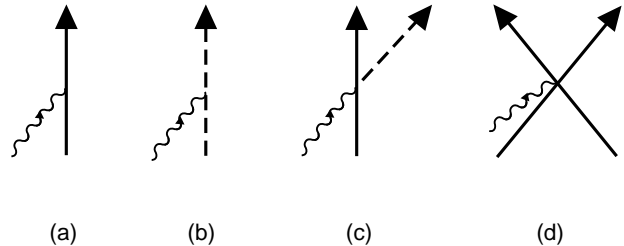,height=1.5in}
\caption{Lagrangian building blocks for EM interactions of pions and 
nucleons. Solid lines are nucleons, dashed lines are pions, and wavy lines
depict photons.}
\end{figure}

Various important building blocks are shown in Fig.\ (6).  The 
impulse-approximation current, $J^{\mu}_{imp}$, is shown in Fig.\ (6a).  The  
relativistic form of the four-current is $\bar{N} \gamma^{\mu} N$,  
and involves no derivatives. We know,  
however, that when ``odd'' $\gamma$-matrices are rendered to  
nonrelativistic form, they are of leading order $(p/M_N)$, while ``even'' ones
are $\cal{O}$(1). Thus, $\Delta_{EM} = -1$ for $\rho_{imp}$ ($f=2, d=0$), while
$\Delta_{EM} = 0$ for ${\bf J}_{imp}$ ($f=2, d=1$). The pion EM current 
in Fig.\ (6b) has a single derivative\cite{21} and $\Delta_{EM} = -1$, 
while the dominant seagull term\cite{20,30} shown in (6c) has
$\Delta_{EM} = -1$ for $\mbox{\boldmath $J$}_{SG}$ and $\Delta_{EM} = 0$ 
for $\rho_{SG}$.  The short-range two-body three-current in (6d) has 
$f = 4, d \geq 1$ and consequently $\Delta_{EM} \geq 1$. This suppresses the
short-range contributions compared to pion-range and impulse-approximation
currents. We note that the $\Delta$-isobar current (assuming 
that we choose not to include the $\Delta$ in our active Hilbert space) 
and the ($\rho \pi \gamma$) current would be (higher-order in 1/$\Lambda$) 
contributions\cite{10} of the form illustrated in Fig.\ (6c), while the  
($\omega \sigma \gamma$) current\cite{10} would have the form in Fig.\ (6d).

We summarize this section by noting that:\\

\indent $\bullet$ The electromagnetic current in a nucleus can be  
treated in a fashion similar to the energy, and this is most conveniently 
done by separating the EM vertex from the strong ones.\\
\indent $\bullet$ Short-range MEC are suppressed compared to pion-range MEC.

\section{\underline{Final Counting of Powers}} 

Before completing the power counting, it is useful to interpret our  
results using a simple, familiar example. The one-pion-exchange amplitude 
in Fig.\ (1a) has the form
$$
V_{\pi} (\mbox{\boldmath $q$}) = -\frac{g^2_A}{f^2_{\pi}}
\frac{\mbox{\boldmath $t$}_1 \cdot \mbox{\boldmath $t$}_2 
\mbox{\boldmath $\sigma$}_1 \cdot \mbox{\boldmath $q$} 
\mbox{\boldmath $\sigma$}_2 \cdot \mbox{\boldmath $q$}}
{\mbox{\boldmath $q$}^2 + m^2_{\pi}} \eqno (8.1)
$$
in momentum space, where $\mbox{\boldmath $q$}$ is the three-momentum  
transferred between nucleons ``1'' and ``2''.  In terms of power
counting, $V_{\pi}({\bf q}) \sim Q^0 / f^2_{\pi}$.  As we argued in the  
previous section and in Appendix A, this should be multiplied by  
the phase-space factor $\frac{d^3 q}{(2 \pi)^3}$ and the Fourier transform
to configuration space completed. The phase-space factor  
behaves as $Q^3$, in agreement with Eq.\ (6.12) (for $n_c = 1, L = 0, 
\Delta = 0$).  Moreover, performing the momentum integrals produces a
Yukawa function multiplied by the familiar factor of $1/4 \pi$ (see Eq.\ 
(3.6)). The latter
converts a single $f_{\pi}$ in Eq.\ (8.1) to a $\Lambda\ (4 \pi f_{\pi} \sim  
\Lambda)$.  Thus $V_{\pi} (r) \sim \frac{Q^3}{\Lambda f_{\pi}}$ as  
we found in Eq.\ (3.8).  We expect that more complicated  
diagrams with more propagators $(n_c > 1)$\cite{39} will generate a  
factor of $1/(4 \pi)^{n_{c}}$.

The interpretation of the short-range force $V_{SR}$ in Fig.\ (3d)  
is somewhat different.  We again have a factor of $(1/f^2_{\pi})$  
from Eq.\ (5.4) and a phase-space factor that leads to a  
$\delta$-function.  We write the latter in the form
$$
\delta^3 ({\bf r}) = \frac{1}{4 \pi} \frac{\delta (r)}{r^2}\, . \eqno (8.2)
$$
The factor of $1/4 \pi$ is the same as before, and $\delta (r) / r^2  
\sim Q^3$, where $Q \sim m_{\pi}$ is the inverse correlation length that 
sets the scale of the correlation function.  Thus, each coordinate-space 
$\delta$-function 
counts as $\sim Q^3/4 \pi$\cite{39}.  This is actually reduced  
somewhat because of the repulsive nature of the short-range correlations.
Nevertheless, combining everything we see that $V_{SR}$ also counts as
$\frac{Q^3}{\Lambda f_{\pi}}$, as predicted by Eq.\ (6.12) with $L = \Delta =  
0$ and $n_c = 1$. 

\begin{figure}[htb]
\epsfig{file=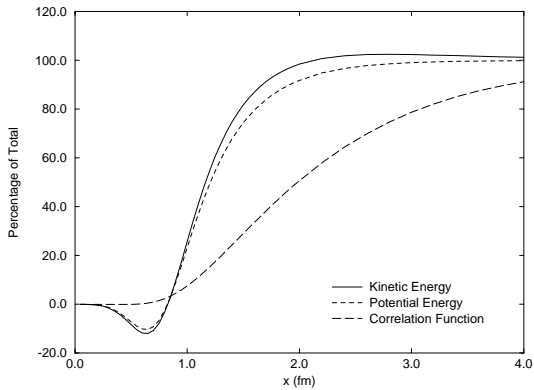,height=3.0in,
angle=-90}
\caption{Percentages of accrual of kinetic energy (solid line),
potential energy (short dashed line), and probability (long dashed line)
within an interparticle separation, $x$, for any pair of nucleons in the
triton. }
\end{figure}

These concepts are illustrated nicely in Fig.\ (7), which shows 
the accrual of potential and kinetic energy in the triton.  A single  
pair of nucleons separated by a distance $x_{12}$ is selected,  
and the expectation value of $\Theta (x - x_{12}) \hat{O}$ is  
calculated, where $\hat{O} = T, V$, or 1.  These values are divided  
by the value for $x = \infty$, and the percentage accrual is  
plotted.  One sees that the major contribution is between 1 and 2  
fm.  Moreover, the short-range part of the potential energy is  
rather modest, starting out repulsive and then yielding to the  
attractive OPEP.  One sees in these plots that $Q \sim m_{\pi}  
\sim [1.4 \, {\rm fm}]^{-1}$ is a reasonable value.

We now put together all of the factors for power counting the energy  
in $D = 4$  dimensions: $Q, f_{\pi}, \Lambda$.  The Lagrangian  
scale factors are given in Eq.\ (5.4).  We also expect a factor of  
$(4 \pi)^{-2L}$ from loops (see Appendix B for a discussion of this and a 
possible counterexample) and $(4 \pi)^{-n_{c}}$  
from configuration-space propagators (see above). This gives our final 
result for $\langle \hat{E} \rangle $, where $\hat{E}$ is any {\it irreducible}
contribution to the energy, in terms of the different scales:
$$
\langle \hat{E} \rangle \sim \frac{Q^{1 + 2(n_{c} + L) + \Delta} f^{2 V - f -  
b}_{\pi}}{\Lambda^{\Delta} (4 \pi)^{2L} (4 \pi)^{n_{c}}} \\ \nonumber
$$
$$
= Q \left( \frac{Q}{f_{\pi}} \right)^{n_{c}} \left( \frac{Q}{\Lambda}  
\right)^{2L + n_{c} + \Delta}, \eqno (8.3)
$$
using $f + b \equiv \sum_i (b_i + f_i ) = 2V + 2(n_c + L)$ obtained  
from Eqs.\ (6.5 - 6.9), and $\Lambda \sim 4 \pi f_{\pi}$.  When counting for 
other observables (viz., the T-matrix), the number of external bosons  
explicitly enters the equations\cite{11}, and this changes the factors 
that set the energy scales.  Our formula reproduces all of  
the previous results obtained using less sophisticated techniques.   
Specific applications are relegated to the next section.

We summarize this section by noting that:\\

\indent $\bullet$ Various nuclear energies behave as $Q^{\nu} /  
\Lambda^{\nu - n_c - 1} f^{n_c}_{\pi}$, with more complicated mechanisms having 
larger values of $\nu$ and $n_c$.\\
\indent $\bullet$ This formula includes phase-space factors required for
conversion to configuration space, and incorporates momentum sharing between 
the nucleons.

\section{\underline{Results and Discussion}}

The most important aspect of this work concerns the relative sizes  
of $N$-nucleon forces.  The leading-order force connecting $N$  
nucleons will have $L = 0, \Delta = 0$, and $n_c = N - 1$,  
corresponding to the simplest possible calculation (all others with  
$L > 0,\, \Delta > 0$ will be smaller).  This produces
$$
\langle V_{N NF} \rangle \sim \frac{Q^{2 N - 1}}{(f_{\pi} \Lambda )^{N - 1}}  
\sim \frac{Q^N}{\Lambda^{N - 1}}\, , \eqno (9.1)
$$
where we have used $Q \sim f_{\pi}$ in order to make the final  
estimate.  We can also use the additional results of Ref.\ \cite{16},  
who found that
$$
\langle V_{NN}\rangle \; \sim 20 \,{\rm MeV/pair}\, , \nonumber 
$$
$$
\langle V_{3NF} \rangle \; \sim 1 \,{\rm MeV/triplet}\, , \nonumber 
$$
$$
\langle V_{4NF}\rangle \; \; \lsim .1 \,{\rm MeV/quartet}\, . \eqno (9.2)
$$
This geometric decrease of the net contribution of many-body  
forces is consistent with Eq.\ (9.1) and is one of the most  
important results of Ref.\ \cite{16}, since it confirms the role of  
chiral symmetry in suppressing many-body forces in nuclei.

Resurrecting the old formalism of Ref.\ \cite{20} (applied to pions  
interacting in a nucleus) also provides some insight into the  
structure of Eq.\ (8.3).  That work was predicated upon performing  
a nonrelativistic expansion of operators using a  
Foldy-Wouthuysen procedure, constructing nuclear operators  
using the superposition principle, and then performing  
time-dependent perturbation theory.  The nucleons in such a  
formalism propagate only forward in time, although mesons go  
forward or backward.  Because the operators refer to the entire  
nucleus (e.g., $J = \sum^A_{i = 1} J_i$  for some impulse-approximation 
vertex, $J_i$), so do the propagators. One simply constructs 
$(E - H)^{-1} \sim [E^{-1}]$ in terms of the nuclear Hamiltonian, $H$  
(details can be found in Refs.\ \cite{19,20}). Thus, power counting  
should be exactly the same as we have already derived, although  
it will be necessary to redefine the variables. This {\it configuration-space} 
formalism automatically incorporates phase-space factors.

\begin{figure}[htbp]
\epsfig{file=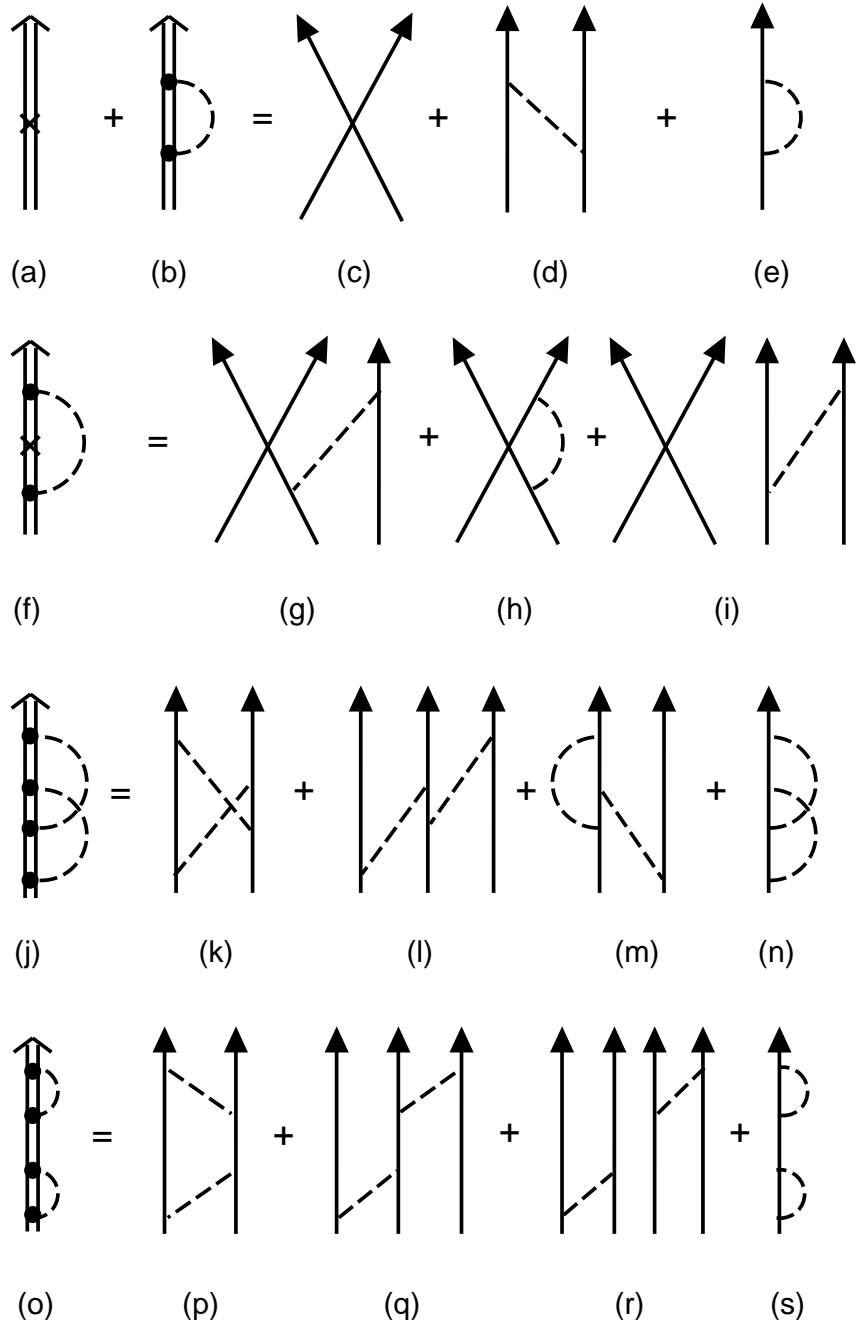,height=7.0in,clip=}
\caption{Contributions to the nuclear energy from ``nucleus''
diagrams on the left, broken down into the usual ``nucleon'' diagrams on the
right. Single solid lines are nucleons, double solid lines represent a nucleus,
dashed lines are pions, while a cross depicts a short-range interaction in 
the nucleus approach.}
\end{figure}

Typical diagrams  
are shown in Figure (8), with diagrams from Ref.\ \cite{20} on the left
expanded as a set of diagrams on the right, where noninteracting
nucleons are suppressed for simplicity. Although the  
``nucleus'' diagrams subsume many distinct mechanisms when  
expanded into ``nucleon'' diagrams, all of the latter share the  
same topology specified by the former, and this we wish to  
explore. The cross represents the short-range interaction, while the  
double lines represent a nucleus propagator, or a nucleus wave  
function, and the large dots represent the pion-nucleus vertex, $J$. Not all 
diagrams are shown.

Figure (8a) for a nucleus is equivalent to Fig.\ (8c).
Figure (8b) can be expanded into Figs.\ (8d) and (8e).  Because the  
vertices in the ``nucleus'' formalism  contain all nucleons,  
expanding second-order perturbation theory leads to both pion  
exchanges (8d) and loops (8e) (only one of the loops is
shown, and this is calculated in Appendix B).  Moreover, we see  
that all possible orderings are included, and that the ``loop-like''  
appearance of the nucleus diagrams results from the  
forward-propagating nucleus.  We count these diagrams as $\nu  
= 1 + 2 \bar{L} + \bar{\Delta}$, where $\bar{L}$ is the number of  
``nucleus loops''.  If we wish to power-count for these ``nucleus''  
diagrams, we must recall that increasing the number of connected
nucleons by 1 increases $\nu$ by 2.  Thus, we must count the  
short-range interaction as $\bar{\Delta} = 2$.  This is a simple  
rearrangement of our original form, moving part of $n_c$ into  
$\Delta$.  Note that if one end of a propagator in a nucleon loop  
(e.g., in Fig.\ (8e)) is detached and reattached to another nucleon  
(as in Fig.\ (8d)), we lower $L$ by 1 and increase $n_c$ by 1,  
keeping $\bar{L} \equiv n_c + L$ fixed, and this is why powers of  
$Q$ depend only on the combination $(n_c + L) = \bar{L}$.  Both  
types count the same and are subsumed in nucleus diagrams  
(e.g., Fig (8b)).  Because different integrals generate different  
factors of $(1 / 4 \pi)$, the counting of $\Lambda$ factors (which leads to 
Eq.\ (8.3); see Appendix B) can be different.  In this example, however, both  
mechanisms behave as $Q^3/\Lambda f_{\pi}$ (in leading order).

Figure (8f) subsumes the graphs of Figs.\
(8g) - (8i), all of which have $\nu = 5$, but differing values of $L,  
n_c, C$, etc.  Note that this set of diagrams includes three-nucleon 
forces, vertex corrections to an $NN$ force (only one of  
which is illustrated in Fig.\ (8h)) and a ``recoil'' graph (with $C = 2$) in  
Fig.\ (8i) (which is a special problem treated later).  The graphs of 
Fig.\ (8j) comprise those of Figs.\ (8k) - (8n) plus several others that are  
disconnected $(C = 2)$, and all have $\nu = 5$.

Our final example treats the ``infrared'' singularities discussed in  
Section 3.  A typical case is Fig.\ (8o), which subsumes
Figs.\ (8p) - (8s).  Nominally the graph has $\bar{L}= 2$ and thus 
$\nu = 5$.  However, if any of the previous graphs are sliced in two by a  
horizontal line, the pion lines are cut, implying a minimum energy  
at that time of $Q \sim E_{\pi}$ in those propagators.  This also holds for 
Fig.\ (8o) as long as the horizontal line intersects a pion propagator.
However, a very different result holds when only the nucleus  
propagator ($\sim \frac{1}{E-H} \sim 1/(Q^2/M_N) \sim \frac{M_N}{Q^2}$,
rather than $1/Q$) is intersected.  In that case  
the energy in the propagator is much smaller.  These diagrams  
are therefore enhanced, as we discussed earlier, when the  
Schr\"{o}dinger equation is solved.  For this reason, one chooses to define 
the Schr\"odinger Hamiltonian in terms of irreducible operators, which  
are then incorporated into the Schr\"{o}dinger equation.  The  
``reducible'' diagrams not only count differently (we must add a factor of 
$\Lambda/Q$ for each infrared propagator, including those in loops, and adjust
factors of $4 \pi$ [see App. C]), but they also contribute to the ``recoil 
graph'' problem.

That problem arises because graphs of the type shown in Fig.\  
(1b) depend on the energy of the nucleus $(E_0)$ and this is  
contained in nuclear propagators:  $(E_{\pi} + H - E_0)^{-1}$.   
Because $H$ and $E_0 \sim Q^2/\Lambda$ and $E_{\pi} \sim  
Q$, it is conventional to expand the propagator in powers of  
$(E_0 - H)/E_{\pi} \sim  \frac{Q}{\Lambda}$.  Although $(E_0 -
H)$ acting on an initial or final wave function vanishes, it can also kill 
the propagator in the middle of Fig.\ (8o), leading to an {\it irreducible}
operator.  Thus the special feature of reducible diagrams leads to  
a ``freedom'' in the form of reducible operators that we choose to
incorporate into our theoretical structure, depending on where we place
parts of various operators. Experience has shown  
that disconnected, but overlapping graphs, of the type shown in Fig.\ (8i) 
can be removed (if desired) using a rearrangement, as can the leading order 
of extended overlapping graphs of the type shown in (8g) and (8l).  This  
rather technical subject can be reviewed in Refs.\ \cite{19,20,24}. The  
practitioner should beware of any graph where a nucleon propagator is not 
required by kinematics to carry an energy $\sim E_{\pi}$ (as in (8g) or  
(8l)) or graphs that are disconnected (as in (8i)).

As an example of how rearrangement affects the power counting,  
it has been shown that the nominal value of $\nu = 5$ (obtained from 
Eq.\ (8.3)) can be changed to $\nu = 6$ (as we already saw in Eq.\ (3.10))
for the 3NF type in Fig.\ (8l), and this also holds for 
the process in Fig.\ (8g).  The short-range 3NF  
resulting from the interaction of 3 nucleons (Fig.\ (3e)) has 
$n_c = 2, \Delta = 1$, and hence $\nu = 6$.  Thus, the leading 3NF can  
be manipulated into $\nu = 6$ (rather than 5) by a suitable definition  
of the nuclear Hamiltonian. Note that there is a factor of $1/\Lambda$
associated with the additional factor of $Q$. These different choices are 
neither right nor wrong; they are a theorist's {\it choice}. One must simply be 
consistent.

\begin{figure}[htb]
\epsfig{file=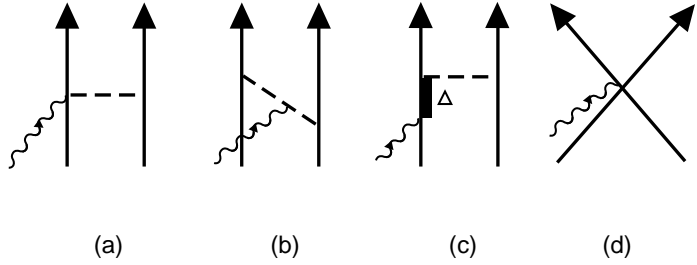,height=1.5in}
\caption{Meson-exchange currents of various types. Solid lines depict 
nucleons, dashed lines show pions, while wavy lines illustrate the EM 
interaction.}
\end{figure}

Our final examples treat a few of the meson-exchange currents in EM  
interactions.  Using the rules we developed earlier, the power  
counting for the four $(n_c = 1)$ graphs in Fig.\ (9) for  
$\vec{J}_{MEC}$ gives 2, 2, 4, 4, respectively, assuming that the  
$\Delta$-isobar is treated as a heavy particle (showing that $\Delta_{EM} \geq 
1$ for Fig.\ (9d) is left as an exercise\cite{10}).  One can show\cite{10,20} 
that graphs (a) and (b) contain a factor of $(4 \pi f^2_{\pi})^{-1}$,  
[(c) and (d) have an additional factor of $\Lambda^{-2}$], so that  
these leading-order MEC behave as $Q^2/f_{\pi} \Lambda \sim  
Q/\Lambda$, and should be comparable to the impulse-approximation result
$\sim \mbox{\boldmath $p$}/M_N \sim Q/ \Lambda$.  This  
precisely conforms to the old ``rules of scale''.  The isobar and  
heavy-meson MEC are suppressed by an additional factor of $(Q/  
\Lambda)^2$.  Moreover, this counting is valid on both sides of  
the current-continuity equation
$$
\mbox{\boldmath $\nabla$} \cdot {\bf J} ({\bf x}) = -i  
[H, \rho ({\bf x})]\, , \eqno (9.3)
$$
using Eq.\ (9.1).

The nonrelativistic charge operator is ${\cal{O}} (1)$, and  
impulse-approximation relativistic corrections are ${\cal{O}} (Q^2/\Lambda^2)$.
The pion-exchange currents (see Appendix A of Ref.\ \cite{19}) are ${\cal{O}} 
(Q^3/4 \pi f^2_{\pi} M_N \sim Q^3/f_{\pi} \Lambda^2 \sim  
Q^2/\Lambda^2)$, consistent with $\Delta_{EM} = 0$ for the seagull  
charge operator.  Note that $Q^2/\Lambda^2$ is really the same  
as $(v/c)^2$, and this also conforms to the old rules of scale\cite{18,20}.

Additional special cases are worked out in the literature\cite{11,12,26,29,30,
38xx}. 

We summarize this section by noting that:\\

\indent $\bullet$ $N$-nucleon forces scale at least as fast as  
$(Q/\Lambda)^{N-1}$, implying that two-nucleon forces are  
stronger than three-nucleon forces are stronger than four-nucleon  
forces \ldots \, .

\indent $\bullet$ Results of recent few-nucleon calculations are  
consistent with this result, which makes nuclear physics tractable.

\indent $\bullet$ The topology of ``nucleus'' (as opposed to  
``nucleon'') diagrams accounts for important aspects of the latter.

\indent $\bullet$ Infrared singularities (reducible diagrams)  
enhance the Schr\"{o}dinger perturbation series.

\indent $\bullet$ Different treatments of ``recoil-graphs'' (resulting from IR  
singularities) can lead to changes in power-counting rules
(always making terms weaker than naive power-counting predictions).

\indent $\bullet$ Power counting for electromagnetic processes is  
consistent with the current-continuity equation.

\indent $\bullet$ Heavy-meson MEC are suppressed relative to  
one-pion-exchange terms.

\section{\underline{Conclusions}}

We have developed systematically the power-counting rules of  
Weinberg, and have added additional expository material and  
numerous examples.  Chiral symmetry provides order, and the  
QCD scale $\Lambda$ plays a deterministic role.  We have  
shown that the nuclear kinetic and potential energies  
(``intrinsic''/pair) scale roughly as $Q^2 / \Lambda$, consistent with  
weakly-bound systems.  $N$-nucleon forces are suppressed as $N$  
increases.  Increasingly complex contributions to the force progressively 
weaken. Short-range meson-exchange currents are weaker than  
pion-range currents, which are comparable to  
impulse-approximation currents.  Large strong-interaction coupling constants 
of heavy-mesons to nucleons result from the mismatch of the scales  
$f_{\pi}$ and $\Lambda$.

All of these results have been stated before in one form or  
another using a variety of arguments or empirical observations,  
but their totality rests on power counting.  In the words of S.\  
Weinberg\cite{42},

\begin{center}
\parbox[t]{3.0in}{``The chiral Lagrangian approach turns out to  
justify assumptions (such as assuming the dominance of  
two-body interactions) that have been used for many years by  
nuclear physicists \ldots''}
\end{center}

\section{\underline{Acknowledgements}}

This work was performed under the auspices of the United States  
Department of Energy. The author would like to thank Bira van Kolck
for providing valuable insights into how strong-interaction physics works and
for advice on the manuscript. He would also like to thank E.\ Tomusiak, 
S.\ Coon, and P.\ Bedaque for similar advice.

\section{\underline{Appendix A -- Momentum Space to  
Configuration Space}}

We have relegated a tedious but instructive part of our derivation  
to this Appendix.  We have chosen the option of power counting  
in configuration space.  A straightforward power-counting  
derivation in momentum space does not involve the last two terms
in Eq.\ (6.10), which arise from the conversion to  
configuration space.  These terms are vital, since they reset the  
baseline (for fixed $A$) against which we determine the importance of  
various operators.  As we will see below, it amounts simply to  
incorporating phase-space factors for each independent degree  
of freedom in a nucleus.  In $D-1$ space dimensions, there are  
$(D - 1)(A - 1)$ independent internal degrees of freedom, plus $D-1$
that specify the motion of the nuclear center-of-mass (CM) and  
don't play any role in our discussion.

We wish to calculate $\langle \Psi_f | \hat{O} | \Psi_i \rangle$ in 
both momentum and configuration spaces (for $D = 4$).  We write
$$
\Psi_{f,i} (\{{\bf r}^{\prime}_{i}\}) = \int \frac{\Pi_i\, (d^3 p^{\prime}_i
\, e^{i {\bf p}^{\prime}_i \cdot {\bf r}^{\prime}_i})}
{(2 \pi )^{3(A - 1)}} \; \; \delta^3 \left(\sum_k {\bf p}^{\prime}_k \right) 
\Psi_{f,i} (\{ {\bf p}^{\prime}_j \})\, , \eqno (A1) 
$$
where our notation ``$\{ \}$'' emphasizes that the wave function  
depends on the {\it set} of internal coordinates $\{ {\bf r}^{\prime}_j \} = 
({\bf r}^\prime_1, {\bf r}^\prime_2, \ldots)$ or momenta 
$\{ {\bf p}^{\prime}_j \}$,
and we have removed the CM coordinates  
($\sum_j {\bf r}^{\prime}_j \equiv 0$ and $\sum_j {\bf p}^\prime_j \equiv 0$).  
It is a convenience to treat  
the coordinates ${\bf r}^{\prime}_j$ as  
independent and use a $\delta$-function as the constraint.  As a  
check, we evaluate the normalization integral:
$$
\langle \Psi_f | \Psi_i \rangle \equiv \int \Pi_i \, ( d^3 r^{\prime}_i)\,
\delta^3 \left(\sum_k {\bf r}^{\prime}_k / A \right) \Psi^\dagger_f 
\left(\{ {\bf r}^{\prime}_j \} \right) \Psi_i (\{ {\bf r}^{\prime}_j \}) \\ 
\nonumber
$$
$$
= \int \frac{\Pi_i (d^3 p^{\prime}_i)}{(2 \pi)^{3 (A - 1)}} \delta^3  
\left(\sum_k {\bf p}^{\prime}_k \right) 
\Psi^\dagger_f ( \{ {\bf p}^{\prime}_j \} )\;
\Psi_i (\{ {\bf p}^{\prime}_j \} ) = 1\; . \eqno (A2)
$$
The factor of $A$ in the ${\bf r}^{\prime}_k$ $\delta$-function is 
conventional.  Thus, in order to obtain the expectation value
of an operator in configuration space, $\langle \hat{O} \rangle$, 
we need to: (1) look at the  
operator in momentum space and (2) multiply by the requisite  
number of independent phase-space factors as given in Eq.\ (A2). This adds 
$[(D - 1)(A - 1)]$ momentum factors to $\nu$ (accounting for the last  
term in Eq.\ (6.10)) and resets the baseline for each diagram or process.  
Note that if we evaluate $\langle \Psi_f | T | \Psi_i \rangle$, where  
$T$ is the kinetic energy, we simply insert $\sum_j  
{\bf p}^{\prime\, 2}_j / 2\, M_N$ between the  
wave functions in the second form of Eq.\ (A2), which is an  
obvious result.

If we take the expectation value of a two-body operator, $V_{ij}  
(r_{ij})$, then a somewhat different form results
$$
\langle V_{12} \rangle = \int \frac{\Pi_i (d^3 q^{\prime}_i) }
{(2 \pi)^{3 (A - 1)}} 
\delta^3 \left( \sum_k {\bf q}^{\prime}_k \right) \int  
\frac{d^3 p}{(2 \pi)^3}  \\ \nonumber
$$
$$
\Psi^\dagger_f ({\bf p} + {\bf q}^{\prime}_1, 
- {\bf p} + {\bf q}^{\prime}_2, \ldots)\, V_{12} ({\bf p})\, \Psi_i
({\bf q}^{\prime}_1, {\bf q}^{\prime}_2, \ldots)\, . \eqno (A3)
$$
In addition to the conventional offset $((D - 1)(A - 1))$, the  
two-body potential inherits a phase-space factor: $\frac{d^3 p}{(2 \pi)^3}$.
This is already included in Eq.\ (6.11) because of momentum  
sharing.  The phase-space factors above serve to kill all the  
$\delta$-functions from non-interacting nucleons, leaving  
phase-space factors (in effect) only for {\it interacting} nucleons.  This  
accounts for the $\frac{d^3 p}{(2 \pi)^3}$ in Eq.\ (A3).  Thus by including 
momentum sharing with ``non-interacting'' nucleons and a full set of  
phase-space factors, we have reset the baseline so that our  
power counting works in the same fashion for any nucleus and  
any operator.  Our final results do not depend on $A$.  Naive power  
counting of the potential in momentum space produces $Q^0$ for OPEP; 
the phase space factor makes  
this $Q^3$, as we derived earlier $(n_c = 1, \Delta = L = 0 \rightarrow \nu  
= 3)$.  Three-nucleon operators pick up an additional  
phase-space factor, and so on.  This completes the interpretation  
of various factors in the power counting.

We summarize this appendix by noting that:\\

\indent $\bullet$ Inclusion of Fourier-transform phase-space factors resets 
the baseline for all diagrams.

\indent $\bullet$ This generates a diagram-independent offset  
$[(D - 1)(A - 1)]$ that makes our final power-counting formula  
independent of $A$.

\section{\underline{Appendix B}}
 
\begin{figure}[htb]
\epsfig{file=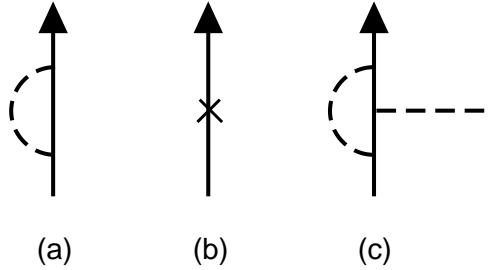,height=1.5in}
\caption{A single nucleon interacting with pions. Solid lines are nucleons, 
dashed lines are pions, and the cross depicts a mass counter term.}
\end{figure}

A wide variety of concepts can be illustrated by working out a  
simple example {\it in toto}.  Figure (10a) illustrates the  
self-energy of a nucleon arising from pionic vacuum fluctuations.   
We perform a nonrelativistic calculation of the S-matrix, noting  
that the nucleon self energy $\sum_N$ is traditionally defined by  
$S = -i \Sigma_N\; [ ( 2 \pi )^4 \delta^4 (P_f - P_i)]$:
$$
\Sigma_N = i \left( \frac{g_A}{f_{\pi}} \right)^2 \int \frac{d^4 k}
{(2 \pi)^4} \; \;  \frac{(\mbox{\boldmath $\sigma$} \cdot  
\mbox{\boldmath $k$})^2\, {\bf t}^2}{(k^2_0 - E^2_{\pi} + i\epsilon)
(-k_0 + E - \frac{({\bf p} + {\bf k})^2}{2 M_N} + i\epsilon)}\, , \eqno (B1)
$$
using the Feynman rules:  $\frac{g_A}{f_{\pi}} t^{\alpha}  
\mbox{\boldmath $\sigma$} \cdot {\bf k}$ for  
creating a pion with isospin component $\alpha$, momentum  $k^\mu$, and 
spin $\mbox{\boldmath $\sigma$}$, which corresponds to $\Delta = 0$.  We have  
included (for now) the complete nucleon propagator (energy $E =  
{\bf p}^2/2\, M_N$, momentum ${\bf p}$), and have defined the pion energy 
by $E^2_\pi = {\bf k}^2 + m^2_\pi$.  Using $(\mbox{\boldmath $\sigma$} \cdot 
{\bf k})^2 = {\bf k}^2, {\bf t}^2 = 3/4$, and noting that there is a  
single pole in the lower $k_0$-plane we obtain
$$
\Sigma_N = -\frac{3}{8} \left( \frac{g_A}{f_{\pi}} \right)^2 \int  
\frac{d^3 k}{(2 \pi)^3}\frac{\hspace{0.150in} {\bf k}^2}
{E_{\pi} \left(E_{\pi} + \frac{({\bf p} + {\bf k})^2 - {\bf p}^2}
{2 M_N} \right)}\, . \eqno (B2)
$$
To leading order in $1/M_N$, we can ignore the kinetic-energy  
terms $\sim 1/M_N$.  We also note that the pion energy sets the  
scale in both of the propagators.  Much of the power counting that we did 
earlier depends {\it explicitly} on this fact. Thus to leading order in  
$(1/M_N)$, we obtain the cubically divergent result
$$
\Sigma_N = -\frac{3}{8} \left( \frac{g_A}{f_{\pi}} \right)^2 \int  
\frac{d^3 k}{(2 \pi)^3} \; \; \frac{{\bf k}^2}{({\bf k}^2 + m^2_{\pi})}\, . 
\eqno (B3)
$$

Interpreting this result requires further work.  First, we must  
regularize the integral and render it finite.  Then we must  
renormalize.  The method of choice is dimensional regularization.   
We convert the 3-dimensional integral to $n$ ``space'' dimensions ($\int  
\frac{d^n k}{(2 \pi)^n}$, with $n$ not necessarily an integer), and  
evaluate it for some $n$ where $\Sigma_N$ is finite, renormalize  
the result, and analytically continue this finite value to $n = 3$.  
The angular integrals plus factors of $2 \pi$ give\cite{43}
$$
\int \frac{d \Omega_n}{(2 \pi)^n} = \frac{2}{\Gamma (n/2)  
(4 \pi)^{n/2}}\, , \eqno (B4)
$$
and we see the origin of factors of $4 \pi$. The remaining integral is\cite{44}
$$
\int^{\infty}_0 \frac{d k\; k^{n + 1}}{(k^2 + m^2_{\pi})} = -
\frac{m^n_{\pi}}{2} \cdot \frac{\pi}{\sin(\frac{n \pi}{2})}\, .  \eqno (B5)
$$
For $n < 0$ these integrals are finite, and their extension to $n =  
3$ (or any odd dimension) is also finite. Combining Eqs.\ (B4) and (B5)
for $n = 3$ produces $m^3_\pi /4 \pi$.

At first sight this is a very  
strange procedure, and the reader is referred to Ref.\ \cite{15} for more  
expert justification.  We note that by adding and subtracting $m^2_{\pi}$ 
in the numerator of the integrand in Eq.\ (B3) we obtain 
$\int \frac{d^3 k}{(2 \pi)^3}$, which has no length or 
energy scale and therefore has no obvious physics associated
with it (so we drop it), plus another term. This process can be repeated, 
producing another scaleless integrand ($-m^2_\pi \int \frac{d^3 k}{(2 \pi)^3
k^2}$) plus $\left[ \frac{m^3_{\pi}}{4 \pi} \right]$, the last part of 
which we obtained previously using Eqs.\ (B4) and (B5).  Thus, we finally 
obtain a well-known result\cite{14}
$$
\Sigma_N = -\frac{3 g^2_A m^3_{\pi}}{3 2 \pi f^2_{\pi}} = - \frac{3  
g^2_A m^3_{\pi}}{8 f_{\pi} (4 \pi f_{\pi})} \sim  
\frac{m^3_{\pi}}{f_{\pi} \Lambda}\, . \eqno (B6)
$$
Manipulations of the type used here to interpret our results should
{\it never} be performed when they introduce singularities at $k=0$ (ours
were finite there).

What is the interpretation of our procedure, and why did we keep  
a single term and argue away the rest?  Figure (10b) also occurs  
as a part of $\Sigma_N$.  This term is simply $M^0_N$, the ``bare''  
nucleon rest mass.  The divergent (but scaleless) integrals can be  
viewed as contributions to $M^0_N$. Since we did not know  
$M^0_N$ anyway, adding terms to it does not introduce a complication.  The  
same is true for $\Sigma_N$ above, but its properties are special.   
Our original methodology was to divide the physics into soft  
(long-range) and hard (short-range) parts.  The scaleless terms  
are hard (divergent), but $\Sigma_N$ is soft.  It is proportional to 
$m^3_\pi \sim Q^3$ in accord with our counting rules $(L = 1, n_c = 0,  
\Delta = 0)$ and is {\it nonanalytic} in $m^2_\pi$.  That is, it has the form  
$\sqrt{m^6_{\pi}} = m^2_{\pi} \sqrt{m^2_{\pi}}$.  Because pion masses must
appear in the Lagrangian as powers of $m^2_{\pi}$, the  
square root is special.  Nonanalytic terms are usually logarithms,  
but not always.  It is a common practice\cite{26,34} to separate out all  
nonanalytic (soft) terms and lump all analytic terms (even finite  
ones) with the so-called ``counter'' terms (hard terms such as $M^0_N$).   
Nevertheless, because the nonanalytic terms have a special form (and are
often large),  
we make the separation:  $M_N \equiv M^0_N + \Sigma_N$.  Note that  
$M_N \sim \Lambda$, but $\Sigma_N \sim m^3_\pi / f_\pi \Lambda \sim 
m^2_\pi / \Lambda$ and is significantly smaller, as we expect on the basis 
of power counting ($L = 1, n_c = 0, \Delta =0 \rightarrow \nu = 3$).

Finally, we note that a single factor of $1/4 \pi$ arose in Eq.\ (B6).  
This is somewhat unusual, as normally loop integrals generate $1/(4 \pi)^2$
or something similar.  A single $(1/4 \pi)$ arises in nonrelativistic cases 
corresponding to odd-dimensional (for $n=3$) integrals, such as Eq.\ (B3).  
In each case the nonanalytic term is an odd polynomial.  In the usual case 
such as Fig.\ (10c) one obtains $1/(4 \pi)^2$ and a logarithm.  A useful  
and instructive exercise is to calculate Fig.\ (10a) covariantly.   
One finds that, indeed, all terms generate an explicit factor of  
$1/(4 \pi)^2$.  If one expands that result for small $m_{\pi} / M_N$,  
one finds
$$
\Sigma^{Rel}_N = \frac{a\, M^3_N + b\, M_N  m^2_{\pi} +c\, m^4_{\pi}/M_N}
{(4 \pi f_{\pi})^2} + \Sigma_N  -\frac{3 g^2_A m^4_{\pi} \log (m_{\pi}/M_N)}
{2 M_N (4 \pi f_{\pi})^2}+ \cdots \, ,  \eqno (B7)
$$
where $a$, $b$, and $c$ are dimensionless, and $a$ and $b$ are divergent. The
$a-c$ terms are analytic functions (polynomials) of $m^2_\pi$ and can be 
incorporated directly into $M^0_N$, as we discussed above. The two scaleless 
integrals that we found above contribute to a and b, respectively. The 
logarithmic term is unique in the expansion.

In performing the expansion in Eq.\ (B7) that produces $\Sigma_N$ (Eq.\ (B6)),  
a factor of $\pi /2$ is generated, leading to a single residual factor  
of $\pi$ in the denominator of Eq.\ (B6). The logarithmic term has a 
dimensionless coefficient of $3 g^2_A/2 \sim 2.4$ and a factor of $1/(4 \pi)^2$,
which was {\it assumed} in our derivation of Eq.\ (8.3) that leads to 
$\Sigma_N \sim Q^3/\Lambda^2$. If we force Eq.\ (B6)
into this form a very large dimensionless coefficient of
$-3 \pi g^2_A/2 \sim -7.5$ results. Sometimes this happens.

Our nonrelativistic calculation reproduced the leading-order  
nonanalytic part of the covariant calculation (and was much easier).
Although the analytic parts of the two calculations are different,
they are not required to be the same and this cannot affect the final results, 
since we do not {\it a priori} know $M^0_N$.

We summarize this appendix with the following observations:\\

\indent $\bullet$ Sensible nonrelativistic field-theory calculations are 
possible.

\indent $\bullet$ Dimensional regularization is an easy way to  
make integrals finite.

\indent $\bullet$ Loops can be rendered into analytic (typically  
``hard'') parts that are incorporated into coupling constants and  
nonanalytic (``soft'') parts, which are kept separate.

\indent $\bullet$ Power counting works for loops.

\indent $\bullet$ The pion mass controls the scale of loop  
propagators.

\indent $\bullet$ Factors of $1/(4 \pi)^2$ (and occasionally $(1/4 \pi)$)
arise from loops. 

\indent $\bullet$ Since $\Lambda \sim 4 \pi f_\pi$ and $Q \sim f_\pi$, counting
powers of $Q/\Lambda$ is not very different from counting powers of 
($1/4 \pi$).

\section{\underline{Appendix C:  Zero-Range Model}}

Another excellent example is the zero-range force\cite{11}.  We write the 
Lagrangian for two identical nonrelativistic nucleons\cite{45} interacting via 
a zero-range force as
$$
{\cal{L}} = N^{\dagger}({\bf x},t) ( i \frac{\partial}{\partial t} 
+ \frac{\mbox{\boldmath $\nabla$}^2}{2 M_N} )N({\bf x},t) - \lambda 
(N^{\dagger}({\bf x},t) N({\bf x},t))^2\, . \eqno (C1)
$$
The form of the scattering amplitude generated by this Lagrangian is a
series of loop diagrams, each order (in $\lambda$) being a product
of loops involving the two nucleons. Performing the $k_0$ integral (part of 
the $d^4 k / (2 \pi)^4$ phase-space factor in each loop) leads directly to 
the Schr\"{o}dinger equation corresponding to an energy $E = k^2/2 \mu$ and
the nuclear (CM) Hamiltonian:${\bf p}^2/2 \mu + 2 \lambda  
\delta^3 (\mbox{\boldmath $r$})$, where $\mbox{\boldmath $r$}$  
is the separation of the nucleons and $\mu$ is the reduced mass.   
We have combined the kinetic energies of the two nucleons so that  
$M_N$ is twice $\mu$; note the combinatorial factor (of two) in front of
the potential. That equation can be written in the form\cite{45} (note that 
$G_0 (r) = -\frac{2 \mu}{4 \pi} \frac{e^{i k r}}{r}$):
$$
\Psi ({\bf r}) = \Psi_0 ({\bf r}) + \int  d^3 r^{\prime}\, G_0 ({\bf r} - 
{\bf r}^{\prime})\, V ({\bf r}^{\prime}) \, \Psi ({\bf r}^{\prime})\, , 
\eqno (C2)
$$ 
where $V({\bf r}^{\prime})$ is the  
($\delta$-function) potential, $\Psi_0 ({\bf r})$ is a 
plane wave, and $G_0 ({\bf z})$ is the Green's function:  
$(E - H + i \epsilon)^{-1}$. Performing the integral, we obtain
$$
\Psi ({\bf r}) = \Psi_0 ({\bf r}) + 2 \lambda\, G_0 ({\bf r})\, \Psi (0)\, , 
\eqno (C3)
$$
the ``zero-range'' solution; setting $r$ to 0 produces an algebraic equation
$$
\Psi (0) = \frac{\Psi_0 (0)}{1 - 2 \lambda\, G_0 (0)}\, , \eqno (C4)
$$
from which we obtain the T-matrix\cite{45} (unitarity implies $Im(T) = 
-2 \mu k \vert T \vert^2 /4 \pi$):
$$
T = \frac{2 \lambda}{1 - 2 \lambda\, G_0 (0)}\, . \eqno (C5)
$$
The Green's function at the origin is (of course) linearly divergent,  
but we can apply the regularization techniques of Appendix B:
$$
G_0 (0) = 2 \mu \int \frac{d^3 p}{(2 \pi)^3} \frac{1}{(k^2 - p^2 + i  
\epsilon)} \eqno (C6)
$$
$$
\rightarrow \frac{- k\, \mu\, i}{2 \pi}\, ,
$$
where again the ``$n$-dimensional'' integral is finite (and is left as an  
exercise, being only slightly different from Eq.\ (B5)).  This  
produces
$$
T = \frac{2 \lambda}{1 + \frac{k\, \mu\, \lambda i}{\pi}}\, , \eqno (C7)
$$
which corresponds to a scattering length, $a = \mu\, \lambda/\pi$, an 
S-matrix,\\
$S = (1-i\, k a)/(1+i\, k a)$,  and an effective-range function
(the inverse of the K-matrix), $k\, \cot(\delta) = -1/a$.

We know that poles in the T-matrix indicate special states.  A  
single pole always exists in this model for
$$
k = \frac{\pi i}{\mu \lambda} \equiv i \kappa\, ,
$$
and
$$
E = \frac{- \kappa^2}{2 \mu} \sim 1/\lambda^2\, . \eqno (C8)
$$
This corresponds to a bound state if $\lambda > 0$ and to a ``virtual''
state if $\lambda < 0$.

This is a most peculiar result, since we started with a potential that  
is {\it repulsive} if $\lambda > 0$, and as $\lambda \rightarrow 0$  
the bound state gets deeper!  One must remember that the  
original problem (interpreted as a quantum mechanics problem)  
has no solution at all.  We have generated a solution by changing the  
problem, redefining $G_0 (0)$ and making it finite, and the peculiar 
properties of (C8) are a reflection of the original (insoluble) problem.   
This redefinition is equivalent to defining how loops (vacuum  
fluctuations) contribute.  Thus, $\lambda$ is really the {\it  
renormalized} coupling constant (of arbitrary sign after renormalization),
and not the ``bare'' one in Eq.\ (C1). How one might interpret and treat
these peculiarities is discussed in Ref. \cite{48}.

What about $\lambda \rightarrow 0$\, ?  This won't happen in  
general, since applying Eq.\ (5.4) leads to the equivalence
$$
\lambda \equiv \frac{c
_{\lambda}}{f^2_{\pi}}\, , \eqno (C9)
$$
with $\vert c_{\lambda}\vert \sim 1$, and using $2 \mu = M_N \sim  
\Lambda \sim 4 \pi f_{\pi}$, we find
$$
\kappa = \frac{(4 \pi f_{\pi})f_\pi }{(2 \mu)2 c_{\lambda}} \sim
\frac{f_\pi}{2 c_\lambda} \sim Q \, , 
\eqno (C10)
$$
as well as $E \sim \frac{- f^2_{\pi}}{4  
c^2_{\lambda} \Lambda}$ and $ a \sim 2 c_\lambda/f_\pi$. A bound state with 
binding energy $\sim$ 2.5 MeV and a corresponding scattering length $\sim 
4.3$ fm is generated for $c_{\lambda} \sim 1$. The case $c_{\lambda} 
\sim - 1$ produces a scattering length $\sim - 4.3$ fm.  We note that an  
anomalously small $c_{\lambda}$ is very improbable.

The unitarity of the T-matrix in lowest-order PT relates the first- and 
second-order amplitudes with a single factor of $1/(4 \pi)$. The (second-order
PT) loop integral must therefore be of the type that generates a {\it single} 
factor, rather than the usual $1/(4 \pi)^2$.

\end{document}